\documentclass[12pt]{article}
\usepackage{epsf}
\def\beq{\begin{equation}}
\def\eeq{\end{equation}}
\def\bea{\begin{eqnarray}}
\def\eea{\end{eqnarray}}

\def\vel{\left|}
\def\ver{\right|}

\def\ga{\left(}
\def\dr{\right)}

\def\rar{\rightarrow}

\def\la{\langle}
\def\ra{\rangle}
\def\ba{\begin{array}}
\def\ea{\end{array}}
\def\ds{\displaystyle}

\title{ {\bf
The exclusive $\bar{B}\rightarrow \pi e^+ e^-$ and 
$\bar{B}\rightarrow \rho e^+ e^-$ decays in the two Higgs doublet model with 
flavor changing neutral currents}}
\author{\vspace{1cm}\\
        {\bf E. O. Iltan}
        \thanks{E-mail address:
        eiltan@heraklit.physics.metu.edu.tr}
 \\
        Physics Department, Middle East Technical University \\
        Ankara, Turkey\\}

\date{}

\begin{document}
\setlength{\baselineskip}{24pt}
\maketitle
\setlength{\baselineskip}{7mm}

\begin{abstract}
We calculate the leading logarithmic QCD corrections to the matrix
element of the decay $b\rightarrow d e^+ e^-$ in the two Higgs doublet 
model with tree level flavor changing currents (model III). We continue  
studying the differential branching ratio and the CP violating 
asymmetry for the exclusive decays $B\rightarrow \pi e^+ e^-$ and 
$B\rightarrow \rho e^+ e^-$ and analysing the dependencies of these 
quantities on the selected model III parameters, $\xi^{U,D}$, 
including the leading logarithmic QCD corrections. Further, we present the 
forward-backward asymmetry of dileptons for the decay
$B\rightarrow \rho e^+ e^-$ and discuss the dependencies to the model III
parameters. We observe that there is a possibility to enhance the branching 
ratios and suppress the CP violating effects for both decays in the framework
of the model III. Therefore, the measurements of these quantities will be an
efficient tool to search the new physics beyond the SM.  
\end{abstract} 
\thispagestyle{empty}
\newpage
\setcounter{page}{1}
\section{Introduction}
Rare B meson decays, induced by flavor changing neutral current (FCNC) 
$b\rightarrow s (d)$ transitions are one of the interesting research area 
to test the Standard model (SM) at loop level. They are informative 
in the determination of the fundamental parameters, such as 
Cabbibo-Kobayashi-Maskawa (CKM) matrix elements, leptonic decay constants,
etc. and useful for establishing the physics beyond the SM, such as 
two Higgs Doublet model (2HDM), Minimal Supersymmetric extension of the SM
(MSSM) \cite{Hewett}, etc. 

Since the SM predicts the large Branching ratio ($Br$), which is measurable 
in the near future, the exclusive decays induced by $b\rightarrow s l^+ l^-$ 
process become attractive. Such transitions has been investigated extensively 
in the SM, 2HDM and MSSM, in the literature  \cite{R4}- \cite{R17}.
For these transitions, the matrix element contains a term includes the
virtual effects of the top quark proportional to $V_{tb}V_{ts}^*$ 
and additional terms describing the $c\bar{c}$ and $u\bar{u}$ loops, 
proportional to $V_{cb}V_{cs}^*$ and $V_{ub}V_{us}^*$ respectively. 
Using the unitarity of CKM matrix, i.e. $V_{ib}V^*_{is}=0,\, i=u,c,t$,
and neglecting the factor $V_{ub}V_{us}^*$ compared to $V_{tb}V^*_{ts}$
and $V_{cb}V^*_{cs}$, it is easy to see that the matrix element involves 
only one independent CKM factor, $V_{tb}V_{ts}^*$. This causes that the 
CP violating effects are suppressed within the SM \cite{aldeilna,du}.
However, for $b\rightarrow d l^+ l^-$ decay, all the CKM factors    
$V_{tb}V_{td}^*$, $V_{cb}V_{cd}^*$ and $V_{ub}V_{ud}^*$ are at the same
order and this leads to a considerable CP violating asymmetry between the
channels induced by the inclusive $b\rightarrow d l^+ l^-$ and 
$\bar{b}\rightarrow \bar{d} l^+ l^-$ decays. These effects have been studied
in the literature  for the inclusive $b\rightarrow d e^+ e^-$ decay, 
in the framework of the SM  \cite{kruger1}.
The difficulties of the experimental investigation of the inclusive decays
stimulate the study of the exclusive decays. However, the theoretical
analysis of the exclusive decays is complicated due to the hadronic form
factors which can be calculated using non-perturbative methods.  
The dispersion formulation of the light-cone constituent quark model 
is one of the method which can be used to calculate the hadronic
matrix elements. In the literature, the form factors for
$b\rightarrow d e^+ e^-$ induced exclusive 
$B\rightarrow (\pi,\rho) e^+ e^- $ decays have been calculated in the 
framework of this method  \cite{melikhov1,melikhov2}.
The CP violation effects for these exclusive decays have been studied 
in the framework of the SM  \cite{kruger2}.

In this work, we present the leading logarithmic (LLog) QCD corrected 
effective Hamiltonian in the 2HDM with flavor changing neutral currents
(model III) for the inclusive $b\rightarrow d e^+ e^-$ decay and calculate
the differential $Br$  of the exclusive $\bar{B}\rightarrow (\pi,\rho) 
e^+ e^-$ process. Further, we study the CP-violation asymmetry ($A_{CP}$) 
and forward-backward asymmetry ($A_{FB}$) of dileptons for the decay 
$\bar{B}\rightarrow \rho e^+ e^-$.

The paper is organized as follows:
In Section 2, we give the LLog QCD corrected Hamiltonian responsible for
the inclusive $b\rightarrow d e^+ e^-$ decay and calculate the matrix element.
In section 3, we present the $Br$ and  $A_{CP}$ of the exclusive 
$\bar{B}\rightarrow \pi e^+ e^-$ decay and analyse the dependencies of 
the $Br$ and  $A_{CP}$ on the couplings $\bar{\xi}_{bb}^{D}$, 
$\bar{\xi}_{tt}^{U}$. 
In section 4, we study the $Br$ ,$A_{CP}$ and $A_{FB}$ of the exclusive 
$\bar{B}\rightarrow \rho e^+ e^-$ decay . Section 5 is devoted to our 
conclusions. In Appendix, we summarize the essential points of the 
model III  and give the explicit forms of some functions we use in our
calculations.
\section{\bf Leading logarithmic improved short-distance contributions in 
the model III for the decay $b\rightarrow d e^+ e^-$ with additional 
long-distance effects }
In this section, we present the LLog QCD corrections to the 
inclusive $b\rightarrow d e^+ e^- $ decay amplitude in the 2HDM with tree 
level neutral currents (model III). 
The LLog QCD corrections to the  
$b\rightarrow d e^+ e^-$ decay amplitude can be calculated 
using the effective theory. In this method, the heavy degrees of freedom,
$t$ quark, $W^{\pm}, H^{\pm}, H_{1}$, and $H_{2}$ bosons, in the present
case, are integrated out. Here $H^{\pm}$ denote 
charged, $H_{1}$ and $H_{2}$ denote neutral Higgs bosons. The procedure is
to match the full theory with the effective low energy theory at the high 
scale $\mu=m_{W}$ and evaluate the Wilson coefficients from $m_{W}$ down 
to the lower scale $\mu\sim O(m_{b})$. In our calculations we choose the 
higher scale as $\mu=m_{W}$ since the current theoretical restrictions 
\cite{gudalil,ciuchini}  show that the charged Higgs mass is enough 
heavy to neglect the running from $m_{H^{\pm}}$ to $m_W$.

The effective Hamiltonian relevant for the decay $b\rightarrow d e^+ e^-$ 
in the model III is
\begin{eqnarray}
{\cal{H}}_{eff}=-4 \frac{G_{F}}{\sqrt{2}} V_{tb} V^{*}_{td} &\{&
\sum_{i=1,..., 12} ( C_{i}(\mu) O_{i}(\mu)+C'_{i}(\mu)
O'_{i}(\mu) )
\nonumber \\ &+& \lambda_{u} \,  \sum_{i=1,2,11,12}
( C_{i}(\mu) (O_{i}(\mu)-O_{i}^{\prime u}(\mu))+
C'_{i}(\mu) (O'_{i}(\mu)-O_{i}^{\prime u}(\mu) ) \}
\label{hamilton}
\end{eqnarray}
where $O^{(')}_{i}$, $O^{u (')}_{i}$, are the operators given in 
eqs.~(\ref{op1}),~(\ref{op2}) and $C^{(')}_{i}$ are the Wilson 
coefficients renormalized at the scale $\mu$. Here the unitarity 
of the Cobayashi-Maskawa matrix (CKM) is used, i.e. 
$V_{tb} V^*_{td}+V_{ub} V^*_{ud}=-V_{cb} V^*_{cd}$ and the parameter 
$\lambda_u$ is defined as: 
\begin{eqnarray}
\lambda_u=\frac{V_{ub} V_{ud}^*}{V_{tb} V_{td}^*} \nonumber
\label{lamdau1}
\end{eqnarray}
Using Wolfenstein parametrization \cite{wolfen}, $\lambda_u$ can be
written as  
\begin{eqnarray}
\lambda_u=\frac{\rho (1-\rho)-\eta^2}{(1-\rho)^2+\eta^2}- 
i \frac{\eta}{(1-\rho)^2+\eta^2}+O(\lambda^2)\nonumber
\label{lamdau2}
\end{eqnarray}
where $\rho$, $\eta$ and $\lambda\sim 0.221$ are Wolfenstein parameters. 
The parameter $\eta$ (and therefore $\lambda_u$) is the reason for the 
$CP$ violation in the SM.

The operator basis is similar to the one used for model III 
(\cite{alil3} and references therein),
which is obtained by replacing $s$-quark by $d$-quark and adding 
new operators, i.e. $O_i^{u (\prime)}\, , i=1,2,11,12$: 
\begin{eqnarray}
 O_1 &=& (\bar{d}_{L \alpha} \gamma_\mu c_{L \beta})
               (\bar{c}_{L \beta} \gamma^\mu b_{L \alpha}), \nonumber   \\
 O_2 &=& (\bar{d}_{L \alpha} \gamma_\mu c_{L \alpha})
               (\bar{c}_{L \beta} \gamma^\mu b_{L \beta}),  \nonumber   \\
 O_1^u &=& (\bar{d}_{L \alpha} \gamma_\mu u_{L \beta})
               (\bar{u}_{L \beta} \gamma^\mu b_{L \alpha}), \nonumber   \\
 O_2^u &=& (\bar{d}_{L \alpha} \gamma_\mu u_{L \alpha})
               (\bar{u}_{L \beta} \gamma^\mu b_{L \beta}),  \nonumber   \\
 O_3 &=& (\bar{d}_{L \alpha} \gamma_\mu b_{L \alpha})
               \sum_{q=u,d,s,c,b}
               (\bar{q}_{L \beta} \gamma^\mu q_{L \beta}),  \nonumber   \\
 O_4 &=& (\bar{d}_{L \alpha} \gamma_\mu b_{L \beta})
                \sum_{q=u,d,s,c,b}
               (\bar{q}_{L \beta} \gamma^\mu q_{L \alpha}),   \nonumber  \\
 O_5 &=& (\bar{d}_{L \alpha} \gamma_\mu b_{L \alpha})
               \sum_{q=u,d,s,c,b}
               (\bar{q}_{R \beta} \gamma^\mu q_{R \beta}),   \nonumber  \\
 O_6 &=& (\bar{d}_{L \alpha} \gamma_\mu b_{L \beta})
                \sum_{q=u,d,s,c,b}
               (\bar{q}_{R \beta} \gamma^\mu q_{R \alpha}),  \nonumber   \\  
 O_7 &=& \frac{e}{16 \pi^2}
          \bar{d}_{\alpha} \sigma_{\mu \nu} (m_b R + m_d L) b_{\alpha}
                {\cal{F}}^{\mu \nu},                             \nonumber  \\
 O_8 &=& \frac{g}{16 \pi^2}
    \bar{d}_{\alpha} T_{\alpha \beta}^a \sigma_{\mu \nu} (m_b R + m_d L)  
          b_{\beta} {\cal{G}}^{a \mu \nu} \nonumber \,\, , \\  
 O_9 &=& \frac{e}{16 \pi^2}
          (\bar{d}_{L \alpha} \gamma_\mu b_{L \alpha})
              (\bar{l} \gamma_\mu l)  \,\, ,    \nonumber    \\
 O_{10} &=& \frac{e}{16 \pi^2}
          (\bar{d}_{L \alpha} \gamma_\mu b_{L \alpha})
              (\bar{l} \gamma_\mu \gamma_{5} l)  \,\, ,    \nonumber  \\
 O_{11} &=& (\bar{d}_{L \alpha} \gamma_\mu c_{L \beta})
               (\bar{c}_{R \beta} \gamma^\mu b_{R \alpha}), \nonumber   \\
 O_{12} &=& (\bar{d}_{L \alpha} \gamma_\mu c_{L \alpha})
(\bar{c}_{R \beta} \gamma^\mu b_{R \beta}), \nonumber \\
 O_{11}^u &=& (\bar{d}_{L \alpha} \gamma_\mu u_{L \beta})
               (\bar{u}_{R \beta} \gamma^\mu b_{R \alpha}), \nonumber   \\
 O_{12}^u &=& (\bar{d}_{L \alpha} \gamma_\mu u_{L \alpha})
(\bar{u}_{R \beta} \gamma^\mu b_{R \beta})\, ,
\label{op1}
\end{eqnarray}
and the second operator set which are flipped chirality partners of the 
first:
\begin{eqnarray}
 O'_1 &=&  (\bar{d}_{R \alpha} \gamma_\mu c_{R \beta})
               (\bar{c}_{R \beta} \gamma^\mu b_{R \alpha}), \nonumber   \\
 O'_2 &=&  (\bar{d}_{R \alpha} \gamma_\mu c_{R \alpha})
               (\bar{c}_{R \beta} \gamma^\mu b_{R \beta}),  \nonumber   \\
 O_1^{\prime u} &=& (\bar{d}_{R \alpha} \gamma_\mu u_{R \beta})
               (\bar{u}_{R \beta} \gamma^\mu b_{R \alpha}), \nonumber   \\
 O_2^{\prime u} &=& (\bar{d}_{R \alpha} \gamma_\mu u_{R \alpha})
               (\bar{u}_{R \beta} \gamma^\mu b_{R \beta}),  \nonumber   \\
 O'_3 &=& (\bar{d}_{R \alpha} \gamma_\mu b_{R \alpha})
               \sum_{q=u,d,s,c,b}
               (\bar{q}_{R \beta} \gamma^\mu q_{R \beta}),  \nonumber   \\
 O'_4 &=& (\bar{d}_{R \alpha} \gamma_\mu b_{R \beta})
                \sum_{q=u,d,s,c,b}
               (\bar{q}_{R \beta} \gamma^\mu q_{R \alpha}),   \nonumber  \\
 O'_5 &=& (\bar{d}_{R \alpha} \gamma_\mu b_{R \alpha})
               \sum_{q=u,d,s,c,b}
               (\bar{q}_{L \beta} \gamma^\mu q_{L \beta}),   \nonumber  \\
 O'_6 &=& (\bar{d}_{R \alpha} \gamma_\mu b_{R \beta})
                \sum_{q=u,d,s,c,b}
               (\bar{q}_{L \beta} \gamma^\mu q_{L \alpha}),  \nonumber   \\  
 O'_7 &=& \frac{e}{16 \pi^2}
          \bar{d}_{\alpha} \sigma_{\mu \nu} (m_b L + m_d R) b_{\alpha}
                {\cal{F}}^{\mu \nu},                             \nonumber       \\
 O'_8 &=& \frac{g}{16 \pi^2}
    \bar{d}_{\alpha} T_{\alpha \beta}^a \sigma_{\mu \nu} (m_b L + m_d R)  
          b_{\beta} {\cal{G}}^{a \mu \nu}, \nonumber \\ 
 O'_9 &=& \frac{e}{16 \pi^2}
          (\bar{d}_{R \alpha} \gamma_\mu b_{R \alpha})
              (\bar{l} \gamma_\mu l)  \,\, ,    \nonumber  \\
 O'_{10} &=& \frac{e}{16 \pi^2}
          (\bar{d}_{R \alpha} \gamma_\mu b_{R \alpha})
              (\bar{l} \gamma_\mu \gamma_{5} l)  \,\, ,    \nonumber \\
 O'_{11} &=& (\bar{d}_{R \alpha} \gamma_\mu c_{R \beta})
               (\bar{c}_{L \beta} \gamma^\mu b_{L \alpha})\,\, , \nonumber   \\
 O'_{12} &=& (\bar{d}_{R \alpha} \gamma_\mu c_{R \alpha})
          (\bar{c}_{L \beta} \gamma^\mu b_{L \beta})\,\, , \nonumber \\
 O_{11}^{\prime u} &=& (\bar{d}_{R \alpha} \gamma_\mu u_{R \beta})
               (\bar{u}_{L \beta} \gamma^\mu b_{L \alpha})\,\, , \nonumber   \\
 O_{12}^{\prime u} &=& (\bar{d}_{R \alpha} \gamma_\mu u_{R \alpha})
               (\bar{u}_{L \beta} \gamma^\mu b_{L \beta})\,\, ,
\label{op2}
\end{eqnarray}
where  
$\alpha$ and $\beta$ are $SU(3)$ colour indices and
${\cal{F}}^{\mu \nu}$ and ${\cal{G}}^{\mu \nu}$
are the field strength tensors of the electromagnetic and strong
interactions, respectively.

The initial values for the first set of operators (eq.(\ref{op1})) 
\cite{Grinstein1,alil3} are 
\begin{eqnarray}
C^{SM}_{1,3,\dots 6,11,12}(m_W)&=&0 \nonumber \, \, , \\
C^{SM}_2(m_W)&=&1 \nonumber \, \, , \\
C_7^{SM}(m_W)&=&\frac{3 x^3-2 x^2}{4(x-1)^4} \ln x+
\frac{-8x^3-5 x^2+7 x}{24 (x-1)^3} \nonumber \, \, , \\
C_8^{SM}(m_W)&=&-\frac{3 x^2}{4(x-1)^4} \ln x+
\frac{-x^3+5 x^2+2 x}{8 (x-1)^3}\nonumber \, \, , \\ 
C_9^{SM}(m_W)&=&-\frac{1}{sin^2\theta_{W}} B(x) +
\frac{1-4 \sin^2 \theta_W}{\sin^2 \theta_W} C(x)-D(x)+\frac{4}{9}, \nonumber \, \, , \\
C_{10}^{SM}(m_W)&=&\frac{1}{\sin^2\theta_W}
(B(x)-C(x))\nonumber \,\, , \\
C^{H}_{1,\dots 6,11,12}(m_W)&=&0 \nonumber \, \, , \\
C_7^{H}(m_W)&=&\frac{1}{m_{t}^2} \,
(\bar{\xi}^{U}_{N,tt}+\bar{\xi}^{U}_{N,tc}
\frac{V_{cd}^{*}}{V_{td}^{*}}) \, (\bar{\xi}^{U}_{N,tt}+\bar{\xi}^{U}_{N,tc}
\frac{V_{cb}}{V_{tb}}) F_{1}(y)\nonumber  \, \, , \\
&+&\frac{1}{m_t m_b} \, (\bar{\xi}^{U}_{N,tt}+\bar{\xi}^{U}_{N,tc}
\frac{V_{cd}^{*}}{V_{td}^{*}}) \, (\bar{\xi}^{D}_{N,bb}+\bar{\xi}^{D}_{N,sb}
\frac{V_{ts}}{V_{tb}}) F_{2}(y)
\nonumber  \, \, , \\
C_8^{H}(m_W)&=&\frac{1}{m_{t}^2} \,
(\bar{\xi}^{U}_{N,tt}+\bar{\xi}^{U}_{N,tc}
\frac{V_{cd}^{*}}{V_{td}^{*}}) \, (\bar{\xi}^{U}_{N,tt}+\bar{\xi}^{U}_{N,tc}
\frac{V_{cb}}{V_{tb}})G_{1}(y)
\nonumber  \, \, , \\
&+&\frac{1}{m_t m_b} \, (\bar{\xi}^{U}_{N,tt}+\bar{\xi}^{U}_{N,tc}
\frac{V_{cd}^{*}}{V_{td}^{*}}) \, (\bar{\xi}^{D}_{N,bb}+\bar{\xi}^{U}_{N,sb}
\frac{V_{ts}}{V_{tb}}) G_{2}(y) \nonumber\, \, , \\
C_9^{H}(m_W)&=&\frac{1}{m_{t}^2} \,
(\bar{\xi}^{U}_{N,tt}+\bar{\xi}^{U}_{N,tc}
\frac{V_{cd}^{*}}{V_{td}^{*}}) \, (\bar{\xi}^{U}_{N,tt}+\bar{\xi}^{U}_{N,tc}
\frac{V_{cb}}{V_{tb}}) H_{1}(y)
\nonumber  \, \, , \\
C_{10}^{H}(m_W)&=&\frac{1}{m_{t}^2} \,
(\bar{\xi}^{U}_{N,tt}+\bar{\xi}^{U}_{N,tc}
\frac{V_{cd}^{*}}{V_{td}^{*}}) \, (\bar{\xi}^{U}_{N,tt}+\bar{\xi}^{U}_{N,tc}
\frac{V_{cb}}{V_{tb}}) L_{1}(y) \, \, , 
\label{CoeffH}
\end{eqnarray}
and for the second set of operators (eq.~(\ref{op2})), 
\begin{eqnarray}
C^{\prime SM}_{1,\dots 12}(m_W)&=&0 \nonumber \, \, , \\
C^{\prime H}_{1,\dots 6,11,12}(m_W)&=&0 \nonumber \, \, , \\
C^{\prime H}_7(m_W)&=&\frac{1}{m_t^2} \,
(\bar{\xi}^{D}_{N,bd}\frac{V_{tb}^{*}}{V_{td}^{*}}+
\bar{\xi}^{D}_{N,sd})
\, (\bar{\xi}^{D}_{N,bb}+\bar{\xi}^{D}_{N,sb}
\frac{V_{ts}}{V_{tb}}) F_{1}(y)
\nonumber  \, \, , \\
&+& \frac{1}{m_t m_b}\, (\bar{\xi}^{D}_{N,bd}\frac{V_{tb}^*}{V_{td}^{*}}
+\bar{\xi}^{D}_{N,sd} \frac{V_{ts}^*}{V_{td}^{*}}) \, 
(\bar{\xi}^{U}_{N,tt}+\bar{\xi}^{U}_{N,tc}
\frac{V_{cb}}{V_{tb}}) F_{2}(y)
\nonumber  \, \, , \\
C^{\prime H}_8 (m_W)&=&\frac{1}{m_t^2} \,
(\bar{\xi}^{D}_{N,bd}\frac{V_{tb}^*}{V_{td}^{*}}+\bar{\xi}^{D}_{N,sd})
\, (\bar{\xi}^{D}_{N,bb}+\bar{\xi}^{D}_{N,sb}
\frac{V_{ts}}{V_{tb}}) G_{1}(y)
\nonumber  \, \, , \\
&+&\frac{1}{m_t m_b} \, (\bar{\xi}^{D}_{N,bd}\frac{V_{tb}^*}{V_{td}^{*}}
+\bar{\xi}^{D}_{N,sd}\frac{V_{ts}^*}{V_{td}^{*}}) \, (\bar{\xi}^{U}_{N,tt}+
\bar{\xi}^{U}_{N,tc}
\frac{V_{cb}}{V_{tb}}) G_{2}(y)
\nonumber \,\, ,\\
C^{\prime H}_9(m_W)&=&\frac{1}{m_t^2} \,
(\bar{\xi}^{D}_{N,bd}\frac{V_{tb}^*}{V_{td}^{*}}+\bar{\xi}^{D}_{N,sd})
\, (\bar{\xi}^{D}_{N,bb}+\bar{\xi}^{D}_{N,sb}
\frac{V_{ts}}{V_{tb}}) H_{1}(y)
\nonumber  \, \, , \\
C^{\prime H}_{10} (m_W)&=&\frac{1}{m_t^2} \,
(\bar{\xi}^{D}_{N,bd}\frac{V_{tb}^*}{V_{td}^{*}}+\bar{\xi}^{D}_{N,sd})
\, (\bar{\xi}^{D}_{N,bb}+\bar{\xi}^{D}_{N,sb}
\frac{V_{ts}}{V_{tb}}) L_{1}(y)
\,\, ,
\label{CoeffH2}
\end{eqnarray}
where $x=m_t^2/m_W^2$ and $y=m_t^2/m_{H^{\pm}}^2$.
In eqs.~(\ref{CoeffH}) and (\ref{CoeffH2}) we used the redefinition
\begin{eqnarray}
\xi^{U,D}=\sqrt{\frac{4 G_{F}}{\sqrt{2}}} \,\, \bar{\xi}^{U,D}\,\, .
\label{ksidefn}
\end{eqnarray}
Here the Wilson coefficients $C_{i}^{SM}(m_{W})$ and $C_{i}^{H}(m_{W})$
denote the SM and the additional charged Higgs contributions
respectively.
The functions $B(x)$, $C(x)$, $D(x)$, $F_{1(2)}(y)$, $G_{1(2)}(y)$, 
$H_{1}(y)$ and $L_{1}(y)$ are given in appendix B.
Note that in the calculations we neglect the contributions due to 
the neutral Higgs bosons since their interactions include negligible 
Yukawa couplings (see \cite{alil2} for details).

Finally, the initial values of the Wilson coefficients in the model III  
(eqs. (\ref{CoeffH})and (\ref{CoeffH2})) are 
\begin{eqnarray}
C^{2HDM}_{1,3,\dots 6,11,12}(m_W)&=&0 \nonumber \, \, , \\
C_2^{2HDM}(m_W)&=&1 \nonumber \, \, , \\
C_7^{2HDM}(m_W)&=&C_7^{SM}(m_W)+C_7^{H}(m_W) \nonumber \, \, , \\
C_8^{2HDM}(m_W)&=&C_8^{SM}(m_W)+C_8^{H}(m_W) \nonumber \, \, , \\ 
C_9^{2HDM}(m_W)&=&C_9^{SM}(m_W)+C_9^{H}(m_W) \nonumber \, \, , \\
C_{10}^{2HDM}(m_W)&=&C_{10}^{SM}(m_W)+C_{10}^{H}(m_W) \nonumber \, \, , \\ 
\nonumber \\ 
C^{\prime 2HDM}_{1,2,3,\dots 6,11,12}(m_W)&=&0 \nonumber \, \, , \\
C_7^{\prime 2HDM}(m_W)&=&C_7^{\prime SM}(m_W)+C_7^{\prime H}(m_W) \nonumber \, \, , \\
C_8^{\prime 2HDM}(m_W)&=&C_8^{\prime SM}(m_W)+C_8^{\prime H}(m_W) \nonumber \, \, , \\ 
C_9^{\prime 2HDM}(m_W)&=&C_9^{\prime SM}(m_W)+C_9^{\prime H}(m_W) \nonumber \, \, , \\
C_{10}^{\prime 2HDM}(m_W)&=&C_{10}^{\prime SM}(m_W)+C_{10}^{\prime H}(m_W) \, \, . 
\label{Coef2HDM}
\end{eqnarray}
These initial values help us calculate the coefficients $C_i^{2HDM}$ and
$C_i^{\prime 2HDM}$ at any lower scale as in the SM  (\cite{alil1}
references therein). 
The $\mu$ scale dependence of the coefficients in the LLog approximation 
can be found in the literature \cite{buras,Grinstein2,misiak,greub}. 
The operators $O_5$, $O_6$, $O_{11}$, $O_{11}^{u}$, $O_{12}$ and 
$O_{12}^{u}$ ( $O'_5$, $O'_6$, $O'_{11}$, $O_{11}^{\prime u}$, $O'_{12}$ 
and $O_{12}^{\prime u}$) give contribution to the leading order matrix 
element of $b\rightarrow s\gamma$ and the magnetic moment type coefficient 
$C_7^{eff}(\mu)$ ($C_7^{\prime eff}(\mu)$) is redefined in the NDR scheme 
as:
\begin{eqnarray}
C_{7}^{eff}(\mu)&=&C_{7}^{2HDM}(\mu)+ Q_d \, 
(C_{5}^{2HDM}(\mu) + N_c \, C_{6}^{2HDM}(\mu))\nonumber \, \, , \\
&+& Q_u\, (\frac{m_c+m_u}{m_b}\, C_{12}^{2HDM}(\mu) + N_c \, 
\frac{m_c+m_u}{m_b}\,C_{11}^{2HDM}(\mu))\nonumber \, \, , \\
C^{\prime eff}_7(\mu)&=& C^{\prime 2HDM}_7(\mu)+Q_{d}\, 
(C^{\prime 2HDM}_5(\mu) + N_c \, C^{\prime 2HDM}_6(\mu))\nonumber \\
&+& Q_u (\frac{m_c+m_u}{m_b}\, C_{12}^{\prime 2HDM}(\mu) + N_c \, 
\frac{m_c+m_u}{m_b}\,C_{11}^{\prime 2HDM}(\mu))\, \, .
\label{C7eff}
\end{eqnarray}
Since $O_2^{(u)}$ ($O_2^{\prime (u)}$) produce dilepton via virtual photon, 
their Wilson coefficient $C_2(\mu)$ ($C^{\prime}_2(\mu)$) and the coefficients 
$C_1(\mu)$, $C_3(\mu)$, ...., $C_6(\mu)$ ($C^{\prime}_2(\mu)$, $C^{\prime}_3
(\mu)$, ..., $C^{\prime}_6(\mu) $) induced by the operator mixing, give 
contributions to $C_9^{eff}(\mu)$ ($C^{\prime eff}_9(\mu)$).
In a more complete analysis, one has to take into account the long-distance
(LD) contributions, produced by real $u\bar{u}$, $d\bar{d}$ and  $c\bar{c}$
intermediate states, i.e. $\rho$, $\omega$ and $\psi^{(i)},\, i=1,... ,6$ 
(Table \ref{Table1}).
These effects can be taken into account
by introducing a Breit-Wigner form of the resonance propogator and it gives
an additional contribution to $C_9^{eff}(\mu)$ \cite{R10,zakharov} 
($C^{\prime eff}_9(\mu)$). Finally the effective coefficients $C_9^{eff}(\mu)$ 
\cite{kruger1, greub} and  $C^{\prime eff}_9(\mu)$ are defined in the NDR 
scheme as:  
\begin{eqnarray} 
C_9^{eff}(\mu)&=& C_9^{2HDM}(\mu) \tilde\eta (\hat s) + 
\left (h(z, \hat s)-\frac{3}{\alpha^2_{em}}\kappa \sum_{V_i=\psi_i}
\frac{\pi \Gamma(V_i\rightarrow ll)m_{V_i}}{q^2-m^2_{V_i}+i m_{V_i}
\Gamma_{V_i}} \right ) \nonumber \\ & &
\left ( 3 C_1(\mu) + C_2(\mu) + 3 C_3(\mu) + C_4(\mu) + 3
C_5(\mu) + C_6(\mu) \right ) \nonumber \\ 
&+&
\lambda_u \Bigg \{ h(z, \hat s)-\frac{3}{\alpha^2_{em}}\kappa 
\sum_{V_i=\psi_i}\frac{\pi \Gamma(V_i\rightarrow ll)m_{V_i}}
{q^2-m^2_{V_i}+i m_{V_i}\Gamma_{V_i}}  \nonumber \\ 
&-& h(0, \hat s)+
\frac{16 \pi^2}{9} \sum_{V_j=\rho,\, \omega}\frac{f_{V_j}^2(q^2)/q^2}
{q^2-m^2_{V_j}+i m_{V_j}\Gamma_{V_j}} \Bigg \} 
\left ( 3 C_1(\mu) + C_2(\mu) \right )  \nonumber \\
&- & \frac{1}{2} h(1, \hat s) \left ( 4 C_3(\mu) + 4 C_4(\mu) + 3
C_5(\mu) + C_6(\mu) \right ) \nonumber \\
&- &  \frac{1}{2} h(0, \hat s) \left( C_3(\mu) + 3 C_4(\mu) \right) +
\frac{2}{9} \left ( 3 C_3(\mu) + C_4(\mu) + 3 C_5(\mu) + C_6(\mu)
\right ) \,\, ,
\label{C9eff}
\end{eqnarray}
and 
\begin{eqnarray} 
C_9^{\prime eff}(\mu)&=& C_9^{\prime 2HDM}(\mu) \tilde\eta(\hat s) + 
\left (h(z, \hat s)-\frac{3}{\alpha^2_{em}}\kappa \sum_{V_i=\psi_i}
\frac{\pi \Gamma(V_i\rightarrow ll)m_{V_i}}{q^2-m^2_{V_i}+i m_{V_i} 
\Gamma_{V_i}} \right ) \nonumber \\ & &
\left ( 3 C'_1(\mu) + C'_2(\mu) + 3 C'_3(\mu) + C'_4(\mu) + 3
C'_5(\mu) + C'_6(\mu) \right ) \nonumber \\ &+&
\lambda_u   \Bigg \{ h(z, \hat s)-\frac{3}{\alpha^2_{em}}\kappa 
\sum_{V_i=\psi_i}\frac{\pi \Gamma(V_i\rightarrow ll)m_{V_i}}{q^2-m^2_{V_i}+
i m_{V_i}\Gamma_{V_i}}  \nonumber \\ &-& h(0, \hat s)+
\frac{16 \pi^2}{9} \sum_{V_j=\rho,\, \omega}\frac{f_{V_j}^2(q^2)/q^2}
{q^2-m^2_{V_j}+i m_{V_j}\Gamma_{V_j}} \Bigg \}
\left ( 3 C'_1(\mu) + C'_2(\mu) \right ) \nonumber \\
&-& \frac{1}{2} h(1, \hat s) \left ( 4 C'_3(\mu) + 4 C'_4(\mu) + 3
C'_5(\mu) + C'_6(\mu) \right ) \nonumber\\
&-&  \frac{1}{2} h(0, \hat s) \left ( C'_3(\mu) + 3 C'_4(\mu) \right ) +
\frac{2}{9} \left ( 3 C'_3(\mu) + C'_4(\mu) + 3 C'_5(\mu) + C'_6(\mu)
\right ) \,  .
\label{C9effp} 
\end{eqnarray}
where $z=\frac{m_c}{m_b}$ and $\hat s=\frac{q^2}{m_b^2}$.
In the above expression, $\tilde\eta(\hat s)$ represents the one gluon
correction to the matrix element $O_9$ with $m_d=0$ \cite{misiak} 
The functions $\tilde\eta(\hat s)$, $\omega(\hat s)$, $h(z, \hat s)$
and $h(0, \hat s)$ are given in appendix C.
In eqs.~(\ref{C9eff}) and ~(\ref{C9effp}), the phenomenological parameter 
$\kappa=2.3$ is chosen to be able to reproduce the correct value of the 
branching ratio  $Br(B\rightarrow J/\psi X\rightarrow X l\bar{l})=  
Br(B\rightarrow J/\psi X)\, Br(J/\psi\rightarrow X l\bar{l})$ \cite{ali}.

In the derivations of $\rho$ and $\omega$ meson resonance effects, 
we used the $q^2$ dependence of the coupling $f_{V_j}$ through the
expression \cite{terasaki} 
\begin{eqnarray}
f_{V_j}(q^2)=f_{V_j}(0)\left (1+\frac{q^2}{P_{V_j}(0)} (P'_{V_j}(0)+
\tilde{P}_{V_j}(q^2)) \right )\,\, ,
\end{eqnarray}
where the coupling $f_{V_j}$ is defined as 
$\,<0|\bar{q}\gamma_{\mu}q|V_j(q^2)|0>=f_{V_j}(q^2) \epsilon_{\mu}$,
$\,P_{V_j}(0)$ and $P'_{V_j}(0)$ are the subtraction constants 
(Table (\ref{Table2})). The function $\tilde{P}_{V_j}(q^2)$ is 
\cite{terasaki}
\begin{eqnarray}
\tilde{P}_{V_j}(q^2)=\frac{1}{16\pi^2 r} \left ( -4-\frac{20}{3} r+
4(1+2 r)(\frac{1-r}{r})^{1/2} Arctan (\frac{r}{1-r})^{1/2} \right )\, ,
\end{eqnarray}
where $r=q^2/4 m_q^2$ and $m_q$ is the mass of the quark which produces 
the meson. This expression is valid in the region $0\leq q^2 \leq 4 m_q^2$.
For the $q^2$ values, $q^2 > 4 m_q^2$, we use the assumption \cite{terasaki} 
$f_{V_j}(q^2)=f_{V_j}(m_{V_j}^2)$ (Table(\ref{Table2})). 

Finally, neglecting the down quark mass, the matrix element for 
$b \rightarrow d e^+ e^-$ decay is obtained as:
\begin{eqnarray}
{\cal M}&=& - \frac{G_F \alpha_{em}}{2\sqrt 2 \pi} V_{tb} V^*_{td} 
\Bigg\{ \left( \, C_9^{eff}(\mu)\,
\bar d \gamma_\mu (1- \gamma_5) b + 
C_9^{\prime eff}(\mu)\, \bar d \gamma_\mu (1+ \gamma_5) b \, \right)
\,\, \bar e \gamma^\mu e \nonumber \\
&+& \left( \, C_{10}(\mu) \, \bar d \gamma_\mu (1- \gamma_5) b+
C'_{10}(\mu)\, \bar d \gamma_\mu (1+ \gamma_5) b \, \right) \,\,
\bar e \gamma^\mu \gamma_5 e   \\
&-& 2 \left( \, C^{eff}_7(\mu)\, \frac{m_b}{q^2}\, 
\bar d i \sigma_{\mu \nu}q^\nu (1+\gamma_5)  b
+C^{\prime eff}_7(\mu)\, \frac{m_b}{q^2}\, 
\bar d i \sigma_{\mu \nu}q^\nu (1-\gamma_5)
b \, \right) \,\, \bar e \gamma^\mu e \Bigg\}~\nonumber .
\label{matr}
\end{eqnarray}
\begin{table}[h]
    \begin{center}
    \begin{tabular}{|c|c|c|}
    \hline
    \hline \hline
    $\psi$         &$m_\psi\,(GeV)$&       $\Gamma(\psi\rightarrow l^+ l^-$)\, (GeV) \\
    \hline \hline    
    $J/\psi$       & $3.097$       &       $5.28\,10^{-6}$ \\
    $\psi^{(2)}$     & $3.686$       &       $2.35\,10^{-6}$ \\
    $\psi^{(3)}$     & $3.770$       &       $2.64\,10^{-7}$ \\
    $\psi^{(4)}$     & $4.040$       &       $7.28\,10^{-7}$ \\
    $\psi^{(5)}$     & $4.160$       &       $7.80\,10^{-7}$ \\
    $\psi^{(6)}$     & $4.420$       &       $4.73\,10^{-7}$ \\ 
    \hline
        \end{tabular}
        \end{center}
\caption{Masses of $\psi$ mesons and decay widths 
$\Gamma(\psi\rightarrow l^+ l^-$) used in the calculations.} 
\label{Table1}
\end{table}

\begin{table}[h]
    \begin{center}
    \begin{tabular}{|c|c|c|c|c|}
    \hline
    \hline \hline
    $  $          &$f_{V}(0)(GeV)$& $f_{V}(m_{V}^2)(GeV)$&
$P_{V}(0)$  &  $P'_{V}(0)$    \\
    \hline \hline    
    $\rho$       & $0.162$       &       $0.17$ & $-0.7498$ &   $-0.0430$ \\
    $\omega$     & $0.166$       &       $0.180$& $-0.7744$  &  $-0.0430$ \\
    \hline
        \end{tabular}
        \end{center}
\caption{The decay couplings and the substraction constants  
for $\rho$ and $\omega$ mesons.}
\label{Table2}
\end{table}

\section{The exclusive $\bar{B}\rightarrow \pi e^+ e^-$ decay} 
\subsection{The formulation}
Now, we continue to present the differential decay rate and CP 
violating asymmetry in the process $\bar{B}\rightarrow \pi e^+ e^-$.
To calculate the decay width, branching ratio, etc., for the exclusive
$\bar{B}\rightarrow \pi e^+ e^-$ decay, we need the matrix elements
$ \la \pi \vel \bar d \gamma_\mu (1\pm \gamma_5) b \ver \bar{B} \ra$, and
$\la \pi \vel \bar d i \sigma_{\mu \nu} q^\nu (1\pm\gamma_5) b \ver \bar{B}
\ra$.
Using the parametrization 
\begin{eqnarray}
<\pi(p_{\pi}| \bar{d}\gamma_{\mu} (1\pm\gamma_5) b|\bar{B}(p_B)>&=&
(2 p_B-q)_{\mu} f_+(q^2)+q_{\mu} f_- (q^2)  \nonumber \,\, , \\
<\pi(p_{\pi}| \bar{d}i\sigma_{\mu\nu} q^{\nu} (1\pm\gamma_5)
b|\bar{B}(p_B)>&=&
-\{ (2 p_B-q)_{\mu} q^2 -(m_B^2-m^2_{\pi}) q_{\mu} \}
v(q^2) \,\, ,
\label{par}
\end{eqnarray}
where $p_{B}$ and $p_{\pi}$ are four momentum vectors of $B$ and $\pi$ 
mesons respectively and  $q=p_B-p_{\pi}$, we get the double differential 
decay rate:  
\begin{eqnarray}
\frac{d \Gamma (\bar{B}\rightarrow \pi e^+ e^-)}{d\sqrt s \,dz} &=& 
\frac{G_F^2 \alpha_{em}^2 m_B^5 \vel V_{tb} V_{td}^* \ver^2\lambda^{1/2}
\sqrt s }{2^{10}\pi^5} \Omega_{\pi}
\label{dddr}
\end{eqnarray}
Here
\begin{eqnarray}
\Omega_{\pi}=\Bigg\{ |(C_9^{eff}+C_9^{\prime eff})f_+(q^2)+
2 (C^{eff}_7+C^{\prime eff}_7) v(q^2) m_b|^2+
|(C_{10}+C'_{10}) f_+(q^2)|^2 \Bigg\} (1-z^2)
\label{etapi}
\end{eqnarray}
and $z=cos \theta$\,, $\theta$ is the angle between the momentum of the  
electron  and that of $B$ meson in the center of mass frame of the lepton 
pair, 
\begin{eqnarray}
\lambda = 1+t^2+s^2 -2 t - 2 s - 2 t s \, ,
\label{lambda}
\end{eqnarray}
where $t=\frac{\ds{m_{\pi}^2}}{\ds{m_B^2}}$ and 
$s=\frac{\ds{q^2}}{\ds{m_B^2}}$.

For the form factors $f_+(q^2)$ and $v(q^2)$, we use the results due to 
the dispersion formulation of the light-cone constituent quark model
\cite{melikhov2}  
\begin{eqnarray}
f_+(q^2)&=&\frac{f_+(0)}{(1-\frac{q^2}{m_{f_+}^2})^{2.35}} \nonumber \,\, , \\
v(q^2)&=&\frac{v(0)}{(1-\frac{q^2}{m_{v}^2})^{2.31}}
\label{form1}
\end{eqnarray}
where $f_+(0)=0.24$, $v(0)=0.05$ and $m_{f_+}=6.71 \,\, GeV$, 
$m_{v}=6.68 \,\, GeV$.

Let us now turn to the CP-violating asymmetry, which is defined as 
\begin{eqnarray}
A_{CP}= \frac{\frac{d \Gamma (\bar{B}\rightarrow \pi e^+ e^-)}{d\sqrt s }-
\frac{d \Gamma (B\rightarrow \bar{\pi} e^+ e^-)}{d\sqrt s }}
{\frac{d \Gamma (\bar{B}\rightarrow \pi e^+ e^-)}{d\sqrt s }+
\frac{d \Gamma (B\rightarrow \bar{\pi} e^+ e^-)}{d\sqrt s }}\,\, .
\label{cpvio}
\end{eqnarray}
The wilson coefficient $C_9^{eff}$ is the origin of the CP violating 
asymmetry since it is a function of 
$\lambda_u=\frac{V_{ub} V_{ud}^*}{V_{tb} V_{td}^*}$.
With the parametrization 
\begin{eqnarray}
C_9^{eff}&=&\xi_1+\lambda_u \xi_2 \nonumber \,\, , \\
C_9^{\prime eff}&=&\xi^{\prime}_1+\lambda_u \xi^{\prime}_2  \,\, ,
\label{c9eff}
\end{eqnarray}
and using eq. (\ref{cpvio})
we get 
\begin{eqnarray}
A_{CP}=-2 Im (\lambda_u)\, \frac{\Delta_{\pi}}{\Omega_{\pi}}\, \lambda \,\,
\label{piAcp}
\end{eqnarray}
where 
\begin{eqnarray}
\Delta_{\pi}= \Bigg \{ Im (\xi_1^{t*} \xi_2^t) f_+(q^2) +
2 m_b Im (\xi_2^t) (C_7^{eff}+C_7^{\prime eff}) \Bigg \} \,
v(q^2) |f_+(q^2)| 
\label{pidelta}
\end{eqnarray}
and 
\begin{eqnarray}
\xi^t_1&=& \xi_1+ \xi_1^{\prime} \nonumber \,\, \\
\xi^t_2&=& \xi_2+ \xi_2^{\prime}
\label{ksi}
\end{eqnarray}
In our numerical analysis we used the input values given in 
Table (\ref{input}).
\begin{table}[h]
        \begin{center}
        \begin{tabular}{|l|l|}
        \hline
        \multicolumn{1}{|c|}{Parameter} & 
                \multicolumn{1}{|c|}{Value}     \\
        \hline \hline
        $m_c$                   & $1.4$ (GeV) \\
        $m_b$                   & $4.8$ (GeV) \\
        $\alpha_{em}^{-1}$      & 129           \\
        $\lambda_t$            & 0.04 \\
        $\Gamma_{tot}(B_d)$    & $3.96 \cdot 10^{-13}$ (GeV)   \\
        $m_{B_d}$              & $5.28$ (GeV) \\
        $m_{\rho}$             & $0.768$ (GeV) \\
        $m_{\pi}$              & $0.139$ (GeV) \\
        $m_{t}$                & $175$ (GeV) \\
        $m_{W}$                & $80.26$ (GeV) \\
        $m_{Z}$                & $91.19$ (GeV) \\
        $\Lambda_{QCD}$             & $0.214$ (GeV) \\
        $\alpha_{s}(m_Z)$             & $0.117$  \\
        $sin\theta_W$             & $\sqrt {0.2325}$  \\
        \hline
        \end{tabular}
        \end{center}
\caption{The values of the input parameters used in the numerical
          calculations.}
\label{input}
\end{table}
\subsection{Discussion}
In this section, we would like to study the $q^2$ dependencies of the differential $Br$,
and $A_{CP}$ of the decay $\bar{B}\rightarrow \pi e^+ e^-$, 
for the selected parameters of the model III
($\bar{\xi}^{U}_{N tt}$,  $\bar{\xi}^{D}_{N bb}$) , 
using the constraints \cite{alil1} coming from the 
$\Delta F=2\, (F=K,D,B)$ mixing ,the $\rho$ parameter \cite{soni2} 
and the measurement by CLEO collaboration \cite{cleo},
\begin{eqnarray}
Br (B\rightarrow X_s\gamma)= (2.32\pm0.07\pm0.35)\, 10^{-4} \,\, .
\label{br2}
\end{eqnarray}
In the calculations, we take $\bar{\xi}_{N tc} << 
\bar{\xi}^{U}_{N tt}, \,\, \bar{\xi}^{D}_{N bb}$ and 
$\bar{\xi}^{D}_{N ij} \sim 0$ where $i$ or $j$ are first or second
generation indices (see \cite{alil2} for details). Under this assumption 
the Wilson coefficients $C^{\prime}_{7}$, $C^{\prime}_{9}$ and  
$C^{\prime}_{10}$ can be neglected compared to unprimed ones and the neutral 
Higgs contributions are suppressed.

\begin{figure}[htb]
\vskip -1.5truein
\centering
\epsfxsize=3.8in
\leavevmode\epsffile{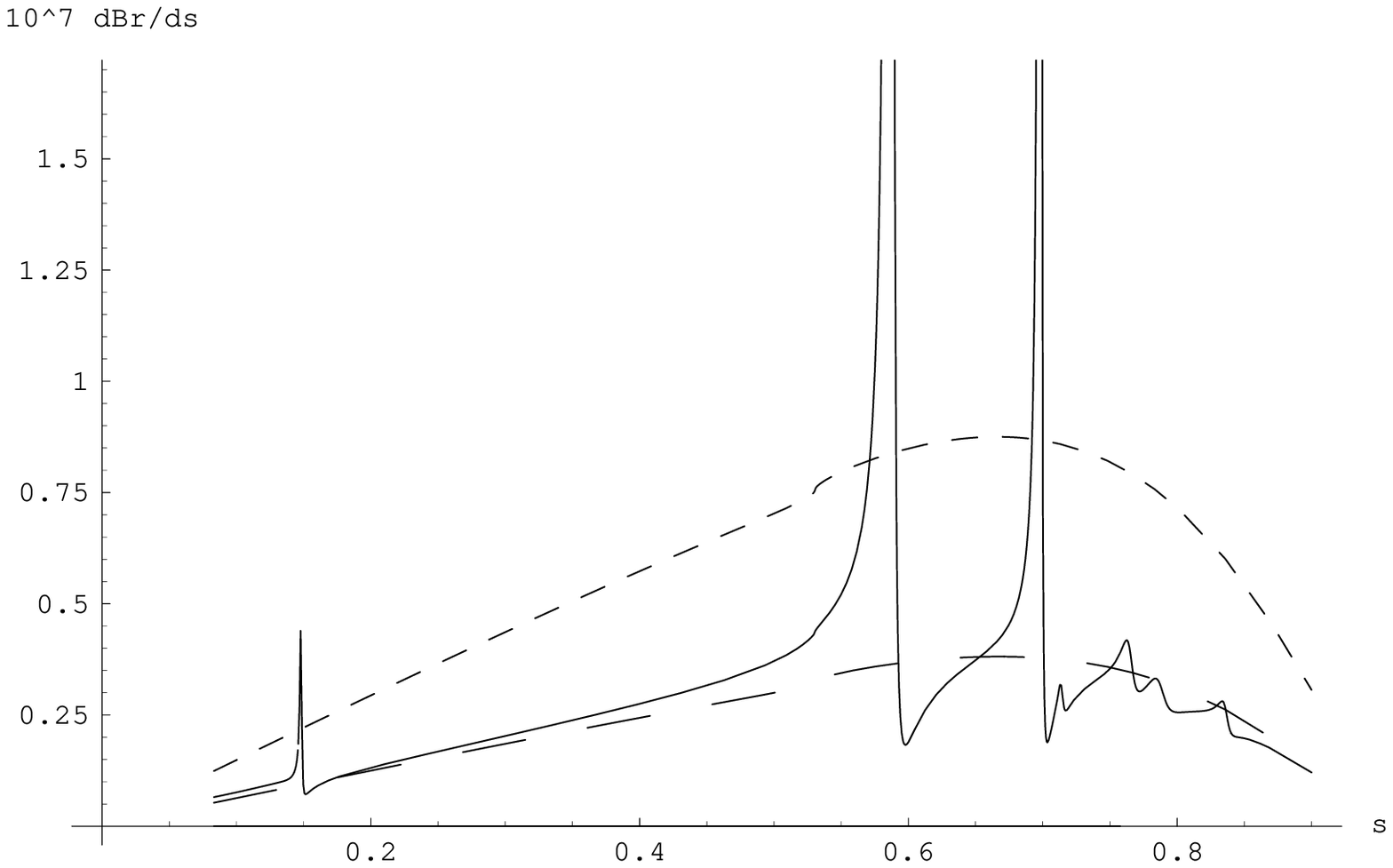}
\vskip -1.5truein
\caption[]{Differential $Br$ as a function of  $q^2$ 
for fixed $\bar{\xi}_{N,bb}^{D}=40\, m_b$ in the region $|r_{tb}|<<1$,
at the scale $\mu=m_b$ for the process $\bar{B}\rightarrow\pi e^+ e^-$.
Here solid line and corresponds to the model III with LD effects,
dashed line to the model III withouth LD effects and dotted dashed line
to the SM withouth LD effects.} 
\label{brbb40q2a}
\end{figure}

\begin{figure}[htb]
\vskip -1.5truein
\centering
\epsfxsize=3.8in
\leavevmode\epsffile{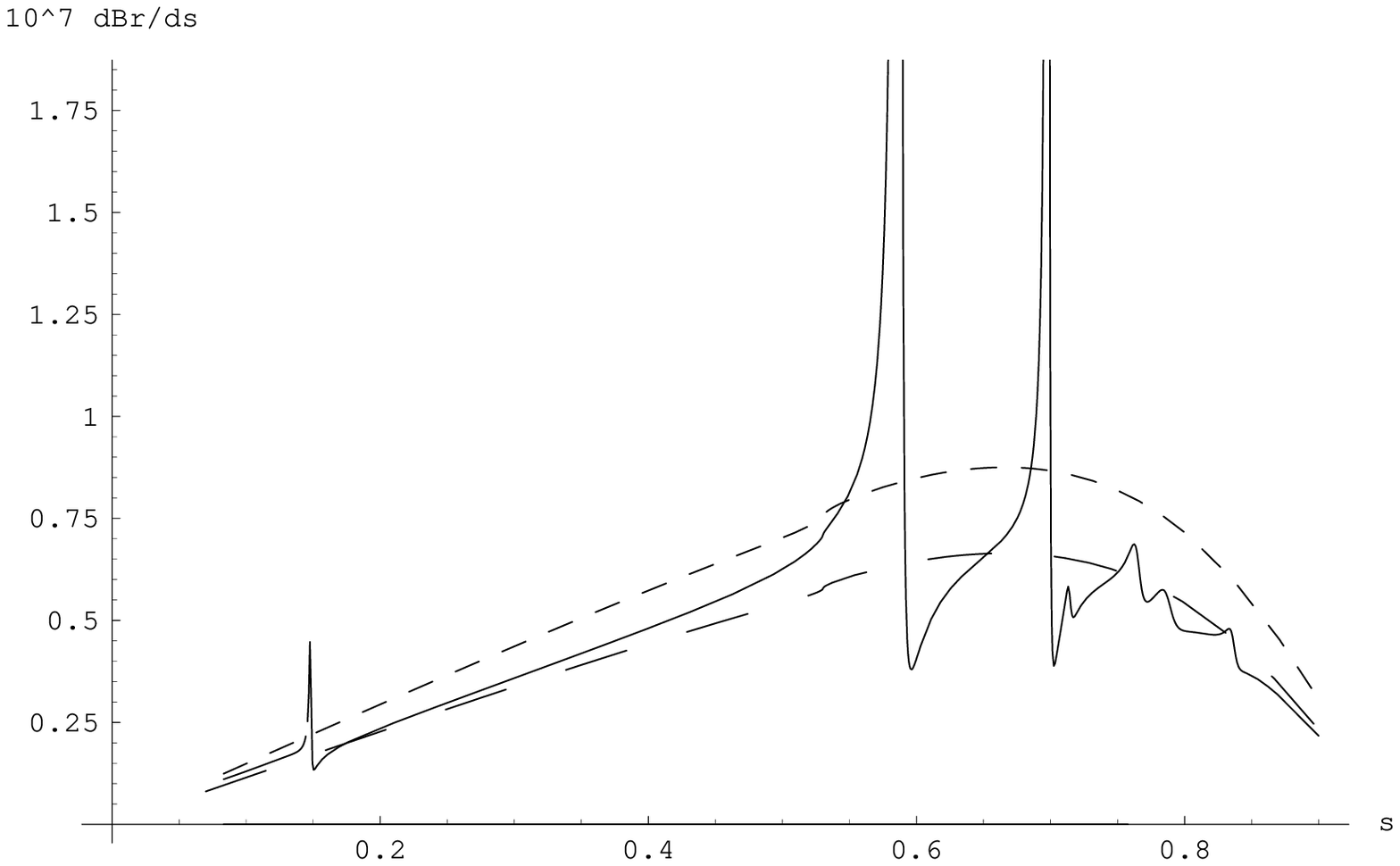}
\vskip -1.5truein
\caption[]{The same as Fig 1, but at the region $r_{tb} >> 1$.}
\label{brbb40q2b}
\end{figure}

\begin{figure}[htb]
\vskip -1.5truein
\centering
\epsfxsize=3.8in
\leavevmode\epsffile{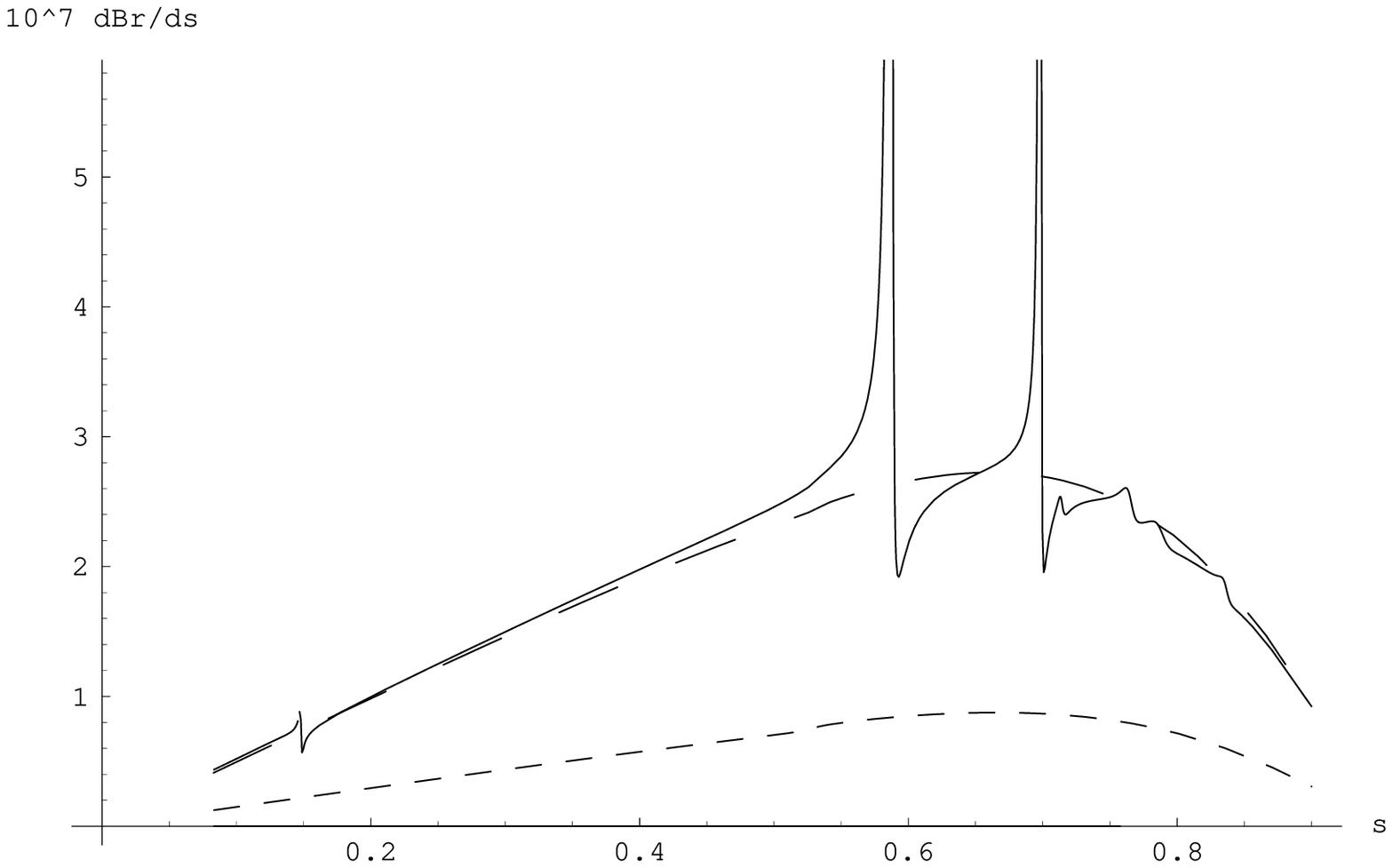}
\vskip -1.5truein
\caption[]{The same as Fig 2, but for fixed $\bar{\xi}_{N,bb}^{D}=90\, m_b$ 
value.}
\label{brbb90q2b}
\end{figure}

In  figs.~\ref{brbb40q2a} and ~\ref{brbb40q2b} we plot the differential 
$Br$ of the decay $\bar{B}\rightarrow  \pi e^+ e^-$ with respect to the 
dilepton mass $q^2$ for the fixed values of $\bar{\xi}_{N,bb}^{D}=40\, m_b$ 
and charged Higgs mass $m_{H^{\pm}}=400\, GeV$
at the scale $\mu=m_b$. Fig.~\ref{brbb40q2a} represents the case where
the ratio $|r_{tb}|=|\frac{\bar{\xi}_{N,tt}^{U}}{\bar{\xi}_{N,bb}^{D}}| <<1.$
It is shown that the differential $Br$ obtained in the model III is smaller
compared to the one calculated in the SM. 
Fig.~\ref{brbb40q2b} (\ref{brbb90q2b}) devoted to the case where $r_{tb}>> 1$ 
for the fixed value of $\bar{\xi}_{N,bb}^{D}$, $\bar{\xi}_{N,bb}^{D}=40 \,
m_b$ ($\bar{\xi}_{N,bb}^{D}=90\, m_b$).  
The differential $Br$ in the model III increases at this region 
($r_{tb}>> 1$) and it enhances strongly compared to the SM with the 
increasing $\bar{\xi}_{N,bb}^{D}$ (Fig.~\ref{brbb90q2b}). 

Now we present the values of $Br$ for the 
$\bar{B} \rar \pi e^+ e^-$ decay in the SM and model III, without LD 
effects. After integrating over $q^2$, we get
\begin{eqnarray}
Br(B \rar \pi e^+ e^-)=0.62 \times10^{-7} \,\,\, (SM)
\label{Br2SM}
\end{eqnarray}
and for the model III
\begin{eqnarray}
Br(B \rar K^* l^+ l^-) = \left\{ \begin{array}{ll}
~ 0.27 \times 10^{-7}   & (|r_{tb}|<<1 \,,\,\bar{\xi}_{N,bb}^{D}=40\, m_b) \\ \\
~ 0.54 \times 10^{-7}   & (r_{tb}>>1 \,,\,\bar{\xi}_{N,bb}^{D}=40\, m_b ) \\ \\
~ 2.65 \times 10^{-7}   & (r_{tb}>>1 \,,\,\bar{\xi}_{N,bb}^{D}=90\, m_b)~.
\end{array} \right.
\label{Br2}
\end{eqnarray}
Here, the strong enhancement of the $Br$ can be observed for 
$r_{tb}>>1$, especially with increasing $\bar{\xi}_{N,bb}^{D}$.  
Note that, in the calculations of $Br$ and the 
differential $Br$, we used the Wolfenstein parameters, 
$\rho=-0.07,\,\, \eta = 0.34$.

\begin{figure}[htb]
\vskip -1.5truein
\centering
\epsfxsize=3.8in
\leavevmode\epsffile{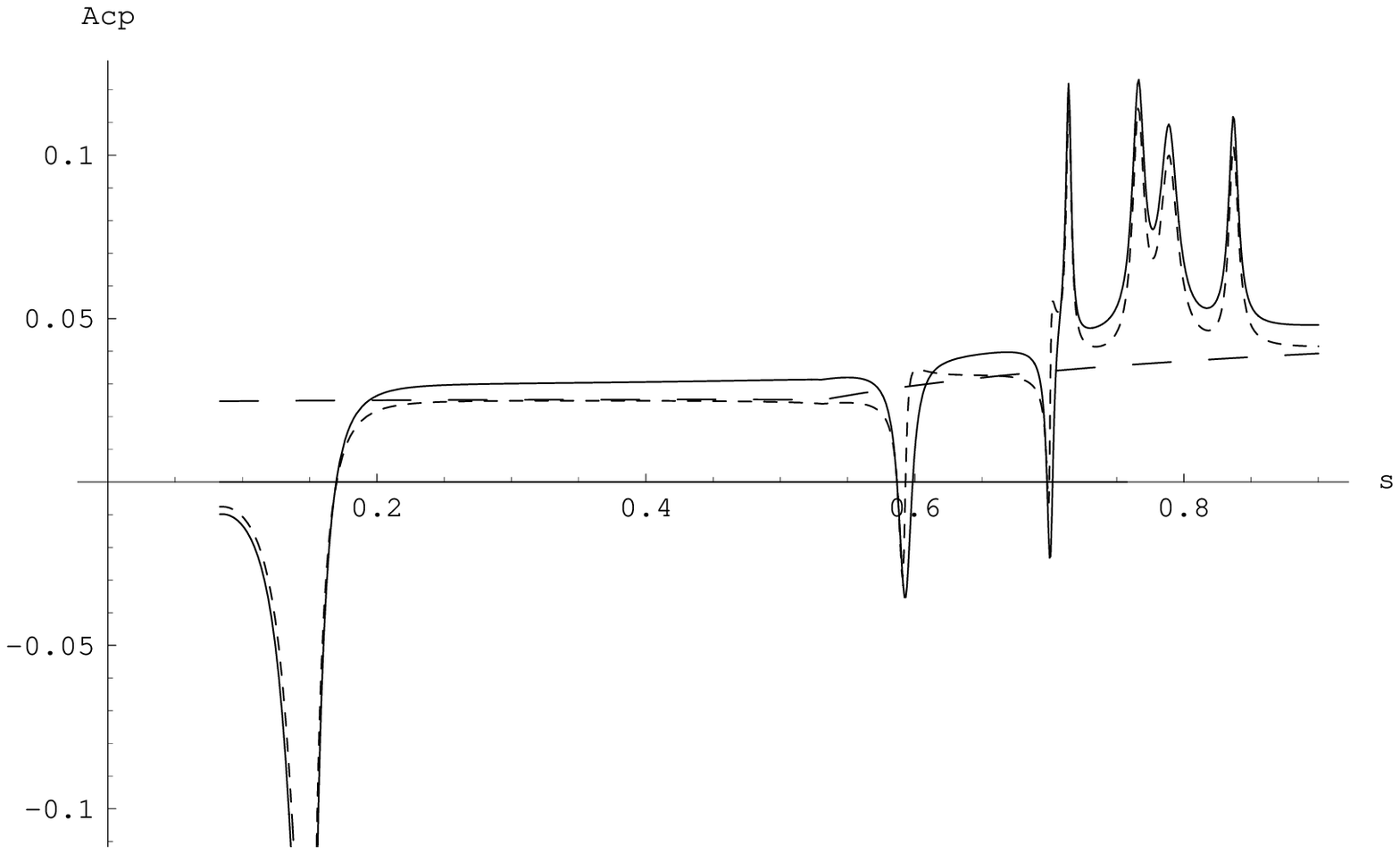}
\vskip -1.5truein
\caption[]{$A_{CP}$ as a function of $q^2$ 
for fixed $\bar{\xi}_{N,bb}^{D}=40\, m_b$ in the region $|r_{tb}|<<1$,
at the scale $\mu=m_b$, for the process $\bar{B}\rightarrow \pi e^+ e^-$.
Here solid line corresponds to the model III with LD effects,
dashed line to the SM withouth LD effects and dotted dashed line
to the SM with LD effects.}
\label{Acpbb40q2a2}
\end{figure}

\begin{figure}[htb]
\vskip -1.5truein
\centering
\epsfxsize=3.8in
\leavevmode\epsffile{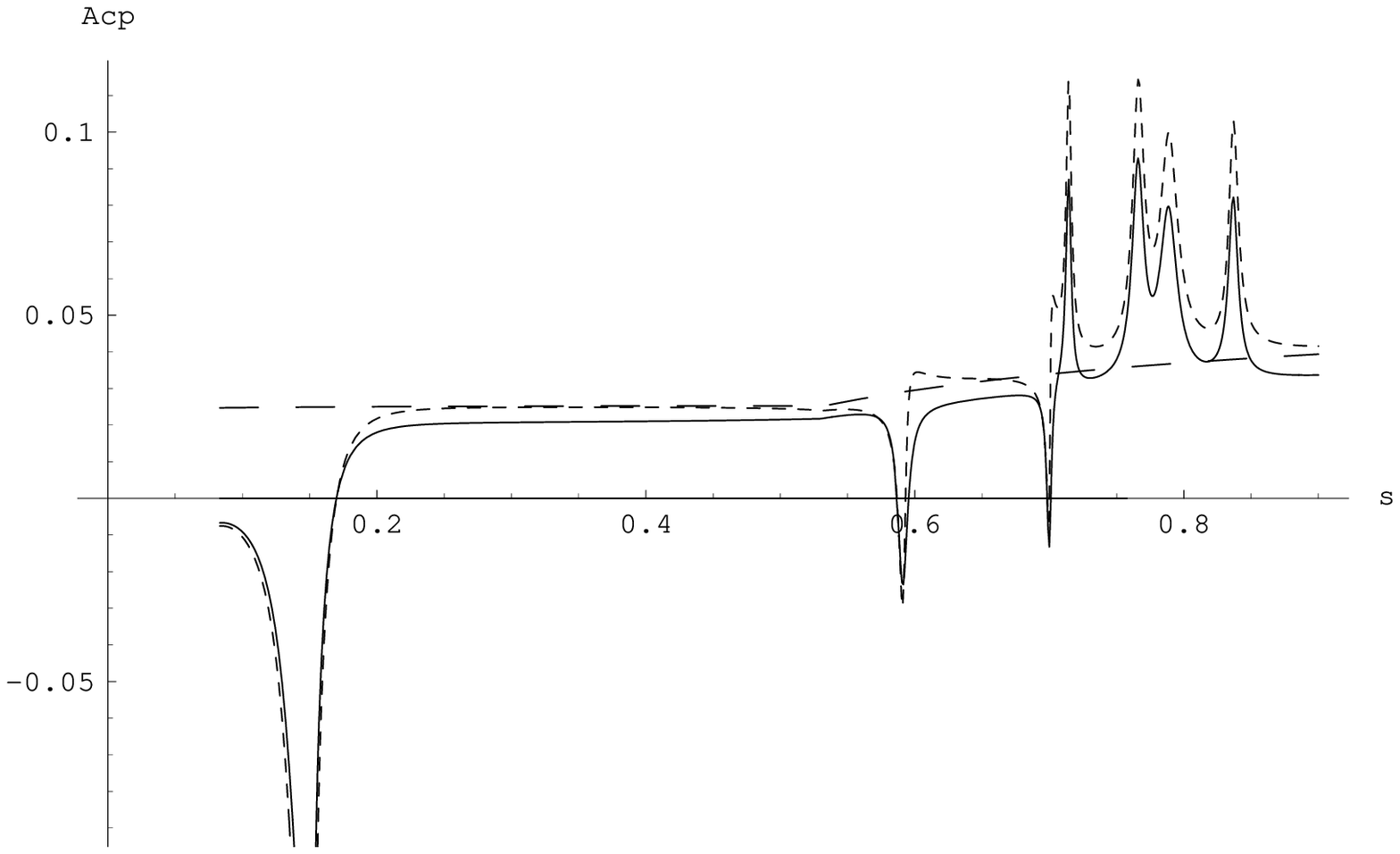}
\vskip -1.5truein
\caption[]{The same as Fig. \ref{Acpbb40q2a2}, 
but at the region $r_{tb} >> 1$ .}
\label{Acpbb40q2b2}
\end{figure}

\begin{figure}[htb]
\vskip -1.5truein
\centering
\epsfxsize=3.8in
\leavevmode\epsffile{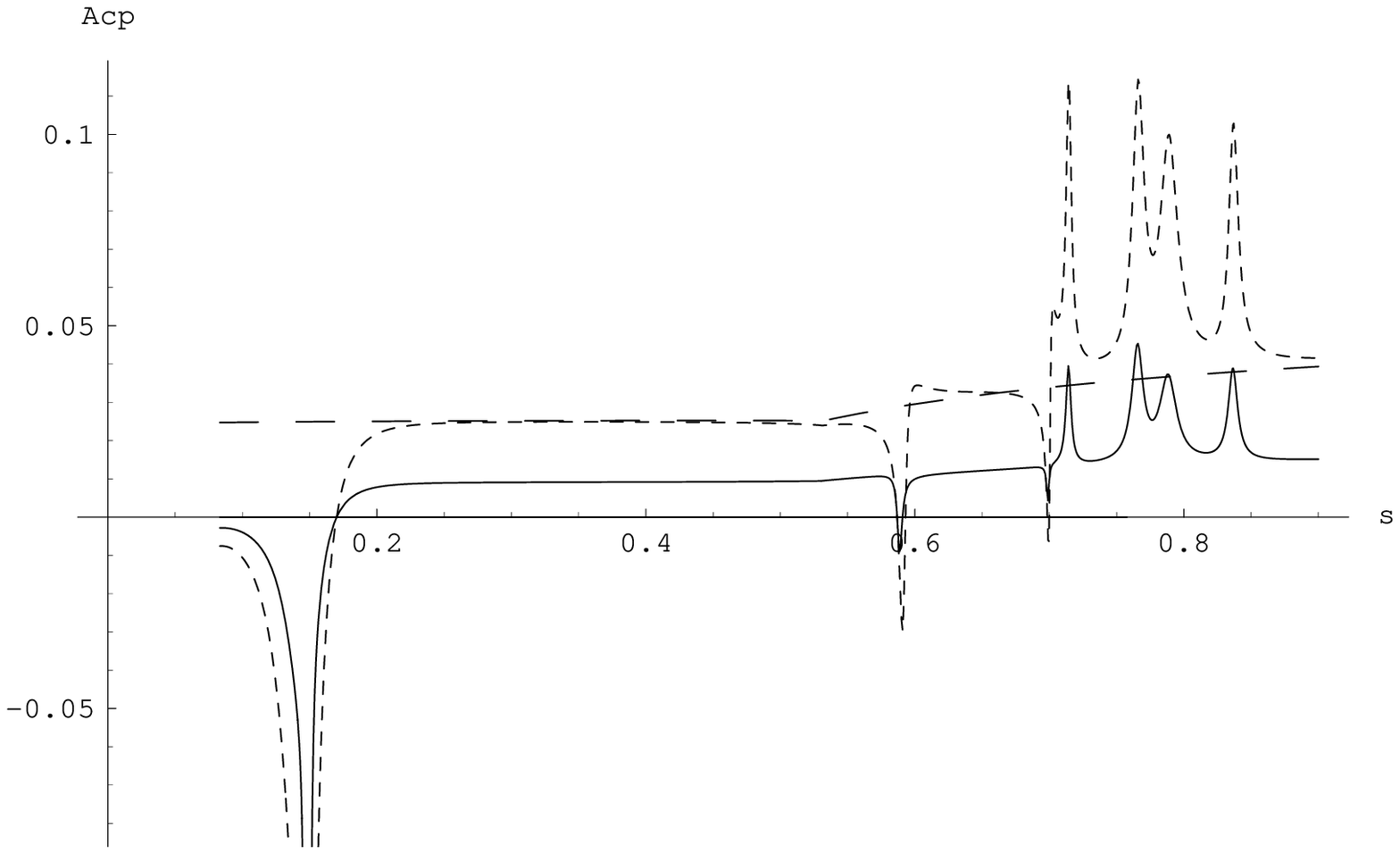}
\vskip -1.5truein
\caption[]{The same as Fig \ref{Acpbb40q2b2}, 
but for fixed $\bar{\xi}_{N,bb}^{D}=90\, m_b$ value. .}
\label{Acpbb90q2b2}
\end{figure}

Figs.~\ref{Acpbb40q2a2} and  \ref{Acpbb40q2b2} show the $q^2$ dependence 
of $A_{CP}$ for the Wolfenstein parameters $\rho=-0.07,\,\, \eta = 0.34$,
fixed values of $\bar{\xi}_{N,bb}^{D}=40\, m_b$ 
and charged Higgs mass $m_{H^{\pm}}=400\, GeV$ at the scale $\mu=m_b$, 
for $|r_{tb}|<< 1$ and $r_{tb}>> 1$ respectively. The CP violation in 
the model III for $|r_{tb}|<< 1$ and $\bar{\xi}_{N,bb}^{D}=40\, m_b$ 
is slightly greater than the one in the SM. However, it decreases for 
$r_{tb}>> 1$ and becomes extremely smaller compared to the one calculated 
in the SM with increasing $\bar{\xi}_{N,bb}^D$ (Fig.~\ref{Acpbb90q2b2}).

We also present $<A_{CP}>$ for two different Wolfenstein parameters in two
different dilepton mass regions

\begin{table}[h]
\small{    \begin{center}
    \begin{tabular}{|c|c|c|c|c|c|}
    \hline
    \hline \hline
    $(\rho,\eta) $   &$SM$&  model III & model III & model III \\
&    &    $\xi_{bb}^D=40 m_b$ & $\xi_{bb}^D=40\, m_b$ & $\xi_{bb}^D=90, 
m_b$&  \\
&    &    $|r_{tb}| <<1$        & $r_{tb}>>1$          & $r_{tb}>>1$& 
$q^2$ regions            \\
\hline \hline
    $(0.3, 0.34)$            &$2.20\, 10^{-2}$ &$2.21\,10^{-2}$
&$1.58\,10^{-2}$ &$0.72\, 10^{-2}$ & I  \\
    \hline
&$0.63 \, 10^{-2}$ &$0.63\,10^{-2}$ &$0.48\,10^{-2}$ &$0.24\, 10^{-2}$& II \\
    \hline \hline

$(-0.07, 0.34)$ &$0.99\, 10^{-2}$ &$1.18\,10^{-2}$ 
&$0.82\,10^{-2}$ &$0.36\, 10^{-2}$& I \\
    \hline
&$0.32 \, 10^{-2}$ &$0.32\,10^{-2}$ &$0.24\,10^{-2}$ &$0.11\, 10^{-2}$& II \\
    \hline \hline    
        \end{tabular}
        \end{center} }
\caption{ The average asymmetry $<A_{CP}>$ for regions I 
( $1\, GeV\leq \sqrt q^2 \leq m_{J/\psi}-20\, MeV$ ) and II 
($m_{J/\psi}+20\, MeV \leq \sqrt q^2 \leq m_{\psi'}-20\, MeV$ ) }
\label{Table3}
\end{table}

In conclusion, we analyse the dependencies of the differential $Br$, $Br$, 
$A_{CP}$ and the average CP-asymmetry $<A_{CP}>$ on the selected model III 
parameters ( $\bar{\xi}_{N,bb}^{D}$, $\bar{\xi}_{N,tt}^{U}$ ) for the decay
$\bar{B}\rightarrow \pi e^+ e^-$. 
We obtain that the strong enhancement of the differential $Br$ ($Br$) is 
possible in the framework of the model III and observe that $A_{CP}$ 
is sensitive to the model III parameters 
($\bar{\xi}_{N,bb}^{D}$,  $\bar{\xi}_{N,tt}^{U}$).

\section{The exclusive $\bar{B}\rightarrow \rho e^+ e^-$ decay} 
\subsection{The formulation}
In this section ,we analyse the differential decay rate, $A_{CP}$ 
and $A_{FB}$ in the process 
$\bar{B}\rightarrow \rho e^+ e^-$. At this stage, we need the matrix elements
$ \la \rho \vel \bar d \gamma_\mu (1\pm \gamma_5) b \ver \bar{B} \ra$, and
$\la \rho \vel \bar d i \sigma_{\mu \nu} q^\nu (1\pm\gamma_5) b \ver \bar{B} 
\ra$. Using the parametrization of the form factors as in \cite{R19}, the 
matrix element of the $\bar{B}\rightarrow \rho e^+ e^-$ decay is obtained 
as \cite{alsav}:
\begin{eqnarray}
{\cal M} &=& -\frac{G \alpha_{em}}{2 \sqrt 2 \pi} V_{tb} V_{td}^*  
\Bigg\{ \bar \ell \gamma^\mu
\ell \left[ 2 A_{tot} \epsilon_{\mu \nu \rho \sigma} \epsilon^{* \nu} 
p_{\rho}^\rho q^\sigma + i
B_{1 \,tot} \epsilon^*_\mu - i B_{2 \,tot} ( \epsilon^* q) 
(p_{B}+p_{\rho})_\mu - 
i B_{3\, tot} (\epsilon^* q)q_\mu \right] \nonumber \\
&+& \bar \ell \gamma^\mu \gamma_5 \ell \left[ 2 C_{tot} \epsilon_{\mu \nu \rho
\sigma}\epsilon^{* \nu} p_{\rho}^\rho q^\sigma + i D_{1\, tot} 
\epsilon^*_\mu - 
i D_{2\, tot} (\epsilon^* q) (p_{B}+p_{\rho})_\mu - i D_{3\, tot} (\epsilon^* q) 
q_\mu \right] \Bigg\}~,
\label{matr2}
\end{eqnarray}
where $\epsilon^{* \mu}$ is the polarization vector of $\rho$ meson, $p_{B}$ 
and $p_{\rho}$ are four momentum vectors of $B$ and $\rho$ mesons, 
$q=p_B-p_{\rho}$ and

\begin{eqnarray}
A_{tot}&=& A+A' \nonumber \,\, , \\
B_{1\, tot}&=& B_1+B'_1 \nonumber \,\, , \\
B_{2\, tot}&=& B_2+B'_2 \nonumber \,\, , \\
B_{3\, tot}&=& B_3+B'_3 \nonumber \,\, , \\
C_{tot}&=& C+C' \nonumber \,\, , \\
D_{1\, tot}&=& D_1+D'_1 \nonumber \,\, , \\
D_{2\, tot}&=& D_2+D'_2 \nonumber \,\, , \\
D_{3\, tot}&=& D_3+D'_3  \,\, .
\label{hadpar0}
\end{eqnarray}
Here
\begin{eqnarray}
A &=& -C_9^{eff} g(q^2)  + 2 C_7^{eff} \frac{m_b}{q^2} g_+(q^2)~,
\nonumber \\ 
B_1 &=& -C_9^{eff} f(q^2)  + 2 C_7^{eff} \frac{m_b}{q^2} 
\left ( (m_B^2 -m_{\rho}^2) g_+(q^2) + q^2 g_-(q^2)\right )~,  \nonumber \\ 
B_2 &=& C_9^{eff} a_+(q^2) + 2 C_7^{eff} \frac{m_b}{q^2} 
\left ( g_+(q^2)+\frac{q^2 h(q^2)}{2}\right )~,  \nonumber \\
B_3 &=& C_9^{eff} a_-(q^2) + 2 C_7^{eff} 
\frac{m_b}{q^2} \left ( g_- (q^2)-\frac{(m_B^2-m_{\rho}^2) h(q^2)}{2}
\right )~,  \nonumber \\ 
C &=& -C_{10}\, g(q^2)~,  \nonumber \\  
D_1 &=& -C_{10}\, f(q^2)~,  \nonumber \\     
D_2 &=& C_{10}\, a_+(q^2)~,  \nonumber \\ 
D_3 &=& C_{10}\, a_-(q^2)~, \nonumber  \\
\label{hadpar1}
\end{eqnarray}
and 
\begin{eqnarray}
A' &=& -C_9^{\prime eff} g(q^2)  + 2 C_7^{\prime eff} \frac{m_b}{q^2} 
g_+(q^2)~, \nonumber\\ 
B'_1 &=& C_9^{\prime eff} f(q^2)  - 2 C_7^{\prime eff} \frac{m_b}{q^2} 
\left ( (m_B^2 -m_{\rho}^2) g_+(q^2) + q^2 g_-(q^2)\right )~,  \nonumber \\ 
B'_2 &=& -C_9^{\prime eff} a_+(q^2) - 2 C_7^{\prime eff} \frac{m_b}{q^2} 
\left ( g(q^2)+\frac{q^2 h(q^2)}{2}\right )~,  \nonumber \\
B'_3 &=& -C_9^{\prime eff} a_-(q^2) - 2 C_7^{\prime eff} 
\frac{m_b}{q^2} \left ( g_- (q^2)-\frac{(m_B^2-m_{\rho}^2) h(q^2)}{2}
\right )~,  \nonumber \\ 
C' &=& -C'_{10}\, g(q^2)~,  \nonumber \\  
D'_1 &=& C'_{10}\, f(q^2)~,  \nonumber \\     
D'_2 &=& -C'_{10}\, a_+(q^2)~,  \nonumber \\ 
D'_3 &=& -C'_{10}\, a_-(q^2)~, \nonumber  \\
\label{hadpar2}
\end{eqnarray}
For the formfactors $g(q^2)$, $a_-(q^2)$, $a_+(q^2)$, $g_+(q^2)$,
$g_-(q^2)$, $h(q^2)$, and $f(q^2)$ we use the dispersion formulation of the
light-cone constituent quark model \cite{melikhov2} in the following 
pole form
\begin{eqnarray}
g(q^2) &=& \frac{0.036}{\displaystyle{\ga 1 - \frac{q^2}{(6.55)^2} 
\dr^{2.75}}}~,~~~
a_+(q^2) = \frac{-0.026}{\displaystyle{\ga 1 - \frac{q^2}{(7.29)^2}
\dr^{3.04}}}~, \nonumber \\
a_-(q^2) &=& \frac{0.03}{\displaystyle{\ga 1 - \frac{q^2}{(6.88)^2}
\dr^{2.85}}}~,~~~
g_+(q^2) = \frac{-0.20}{\displaystyle{\ga 1 - \frac{q^2}{(6.57)^2} 
\dr^{2.76}}}~, \nonumber \\ 
g_-(q^2) &=& \frac{0.18}{\displaystyle{\ga 1 - \frac{q^2}{(6.50)^2} 
\dr^{2.73}}}~,~~~ 
h(q^2) = \frac{0.003}{\displaystyle{\ga 1 - \frac{q^2}{(6.43)^2} 
\dr^{3.42}}}~,  \nonumber \\ 
f(q^2) &=& \frac{1.10}{\displaystyle{\ga 1 - \frac{q^2}{(5.59)^2}+
(\frac{q^2}{(7.10)^2})^2 \dr}}~, 
\label{pole}
\end{eqnarray}
Using eq.(\ref{matr2}), we get the double differential decay rate:  
\begin{eqnarray}
\frac{d \Gamma}{d q^2 dz} &=& \frac{G^2 \alpha_{em}^2
\vel V_{tb} V_{ts}^* \ver^2\lambda^{1/2} }{2^{12}
\pi^5 m_B} \Bigg\{ 2 \lambda m_B^4 \Bigg[ 
m_B^2 s ( 1+ z^2) \ga \vel A_{tot} \ver ^2 +\vel C_{tot} \ver ^2 \dr
\Bigg] \nonumber \\
&+& \frac{\lambda m_B^4}{2 r} \Bigg[ \lambda m_B^2 (1-z^2) \ga \vel 
B_{2\, tot} \ver ^2 + \vel D_{2\, tot} \ver ^2 \dr \Bigg]  \nonumber \\
&+& \frac{1}{2 r} \Bigg[ m_B^2 \left\{ \lambda (1- z^2) + 8 r s\right\}
\ga \vel B_{1\, tot} \ver^2 + \vel D_{1\, tot} \ver^2 \dr  
\nonumber \\
&-& 2 \lambda m_B^4
(1-r-s)(1-z^2)  \left\{ Re\ga  B_{1\, tot} B_{2\, tot}^*  \dr + 
Re \ga D_{1\, tot} D_{2\, tot}^*  \dr \right\} 
\Bigg]  \nonumber \\
&-& 8 m_B^4 s\lambda^{1/2} z \Bigg[ 
\left\{ Re\ga  B_{1\, tot} C_{tot}^* \dr + Re\ga A_{tot} D_{1\, tot}^*\dr 
\right\} \Bigg] \Bigg\}~,
\label{rodddr}
\end{eqnarray}
where $z=cos \theta$\,, $\theta$ is the angle between the momentum of
electron $e$ and that of $B$ meson in the center of mass frame of the 
lepton pair, $\lambda = 1+t^2+s^2 -2 t - 2 s - 2 t s$, $t =
\frac{\ds{m_{\rho}^2}}{\ds{m_B^2}}$ and 
$s=\frac{\ds{q^2}}{\ds{m_B^2}}$.

We continue to present the CP-violating asymmetry, which is defined as 
in eq. (\ref{cpvio}) with the replacement of $\pi\rightarrow\rho$. 
Using the same parametrization as in eq.~(\ref{c9eff}) we get 
\begin{eqnarray}
A_{CP}=-2 Im (\lambda_u)\, \frac{\Delta_{\rho}}{\Omega_{\rho}}\, \lambda \,\,
\label{cpro}
\end{eqnarray}
where 
\begin{eqnarray}
\Delta_{\rho}&=& Im (\xi_1^{t *} \xi_2^{t}) 
\Bigg \{ 4\, s\, m_B^2 \,g^2(q^2)+
\frac{f^2 (q^2)}{\lambda \, m_B^2} (6s + \frac{\lambda}{2 t}) 
+\frac{m_B^2 \lambda}{2 t}\, a^2_+(q^2)
+\frac{(1-s-t)}{t} f(q^2)\, a_+(q^2) \Bigg \}\nonumber \,\, \\
&+& \frac{2 C_7^{eff}}{s} Im(\xi_2)\Bigg \{ -4 \frac{(1+\sqrt t)}
{\sqrt t}\, m_b \, s\, g(q^2)\, g_+(q^2) -\frac{m_b}{2 m_B} 
\left ( (1-t)g_+(q^2)+s\, g_-(q^2)\right )(1+\sqrt t) 
\nonumber \\ & & 
\left (\frac{2 \,f(q^2)}{\lambda m_B}\,(6 s+\frac{\lambda}{2 t})
+ m_B \, a_+(q^2)\, \frac{1-t-s}{t}\right ) 
+\frac{m_b}{2\,m_B\,t} \left ( m_B\, \lambda \, a_+(q^2) + \frac{f(q^2)}{m_B}
(1-t-s)\right ) \nonumber \\ & &
\left (g_+ (q^2)+\frac{s\,m_B^2}{2}\, h(q^2)  \right ) 
\Bigg \}
\label{deltaro}
\end{eqnarray}
and 
\begin{eqnarray}
\Omega_{\rho}&=& \lambda \{ 4 m_{B}^2 s (|A_{tot}|^2+|C|_{tot}^2)+
\frac{1}{m^2_{B} 
\lambda} ( 6 s +\frac{\lambda}{2 t} ) (|B_{1 tot}|^2+|D_{1 tot}|^2) 
\nonumber \\
&+& \frac{\lambda}{2 t} m^2_{B} (|B_{2 tot}|^2+|D_{2 tot}|^2)-
\lambda\frac{1-t-s}{t}
Re (B_{1 tot} B_{2 tot}^* +D_{1 tot} D_{2 tot}^*) \} \,\, .
\label{omegaro}
\end{eqnarray}
Finally, we present $A_{FB}$ which can give more precise information about 
the Wilson coefficients $C_7^{eff},~C_9^{eff}$ and $C_{10}$. 
It is defined as:
\begin{eqnarray}
A_{FB} (q^2) = \frac{\displaystyle{\int_0^1 dz \frac{d \Gamma}{dq^2 dz} - 
\int_{-1}^0dz \frac{d \Gamma}{dq^2 dz}}}{\displaystyle{\int_0^1 dz 
\frac{d \Gamma}{dq^2 dz}+\int_{-1}^0dz\frac{d \Gamma}{dq^2 dz}}}
\label{AFB}
\end{eqnarray}
After the standard calculation, we get 
\begin{eqnarray}
A_{FB}&=& 12\, \lambda^{1/2} \frac{Re(C_{10}+C'_{10})}{\Omega_{\rho}} 
\Bigg \{ s\, f(q^2) \, g(q^2)\,
Re(C_9^{eff}+C_9^{\prime eff})-\frac{m_b}{m_B}
\, (C_7^{eff}+C_7^{\prime eff})\nonumber \\ & &
\left (m_B\, (1+\sqrt t)\, g(q^2)\, 
\left ( (1-t) g_+(q^2) + s \,g_-(q^2) \right )  
+ g_+ (q^2)\, f(q^2) \frac{1+t}{m_B \, (1+\sqrt t)} \right ) \Bigg \}
\label{AFB2}
\end{eqnarray}
\subsection{Discussion}
In this section, we study the $q^2$ dependencies of the differential $Br$,
$A_{CP}$ and $A_{FB}$ of the decay $\bar{B}\rightarrow \rho e^+ e^-$ 
for the selected parameters of the model III
($\bar{\xi}^{U}_{N tt}$,  $\bar{\xi}^{D}_{N bb}$). 
In the calculations, we use the same restrictions for the model III
parameters. (see section 3)

\begin{figure}[htb]
\vskip -1.5truein
\centering
\epsfxsize=3.8in
\leavevmode\epsffile{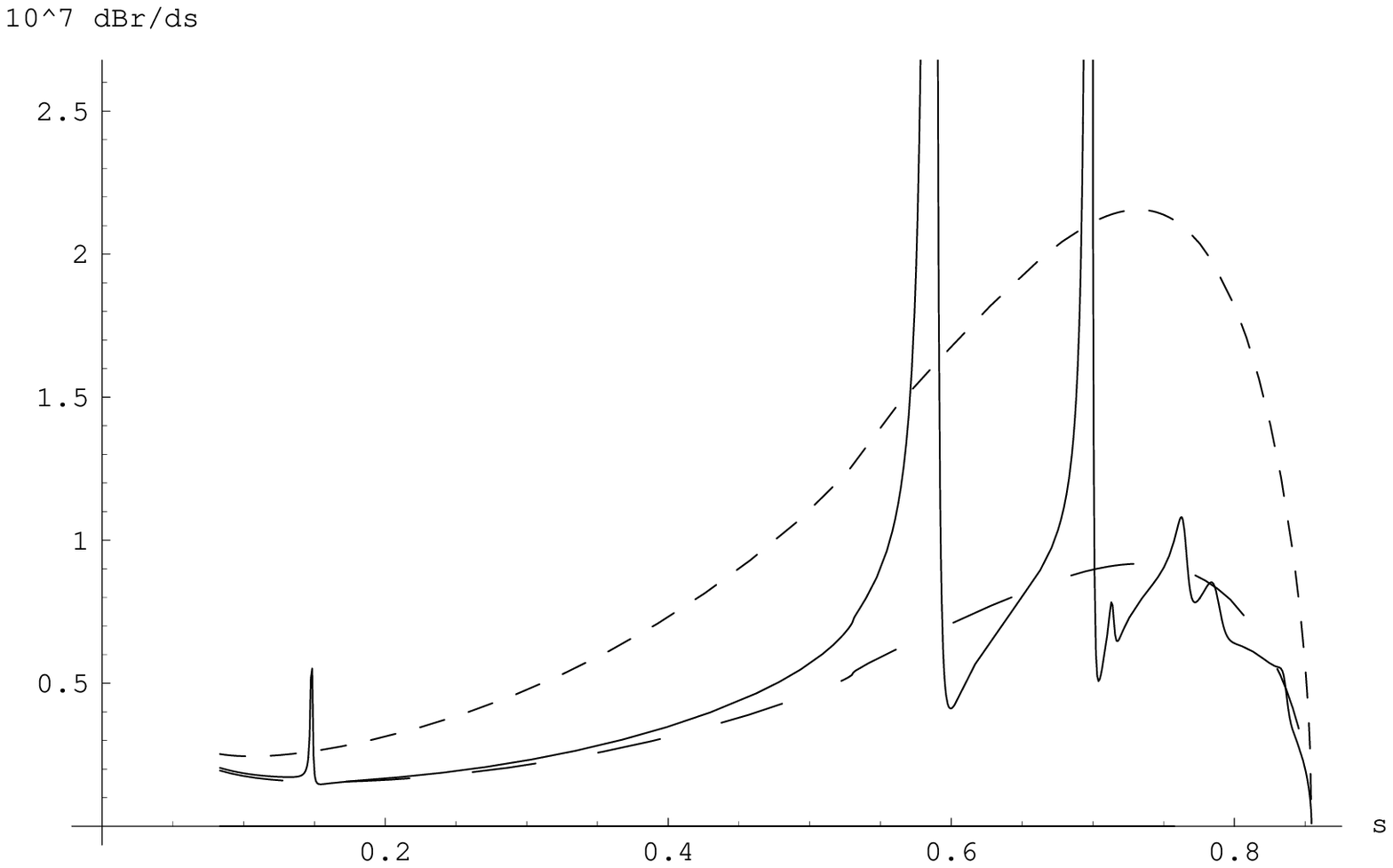}
\vskip -1.5truein
\caption[]{Differential $Br$ as a function of  $q^2$ 
for fixed $\bar{\xi}_{N,bb}^{D}=40\, m_b$ in the region $|r_{tb}|<<1$,
at the scale $\mu=m_b$ for the process $\bar{B}\rightarrow \rho e^+ e^-$.
Here solid line and corresponds to the model III with LD effects,
dashed line to the model III withouth LD effects and dotted dashed line
to the SM withouth LD effects.} 
\label{robrbb40q2a}
\end{figure}

\begin{figure}[htb]
\vskip -1.5truein
\centering
\epsfxsize=3.8in
\leavevmode\epsffile{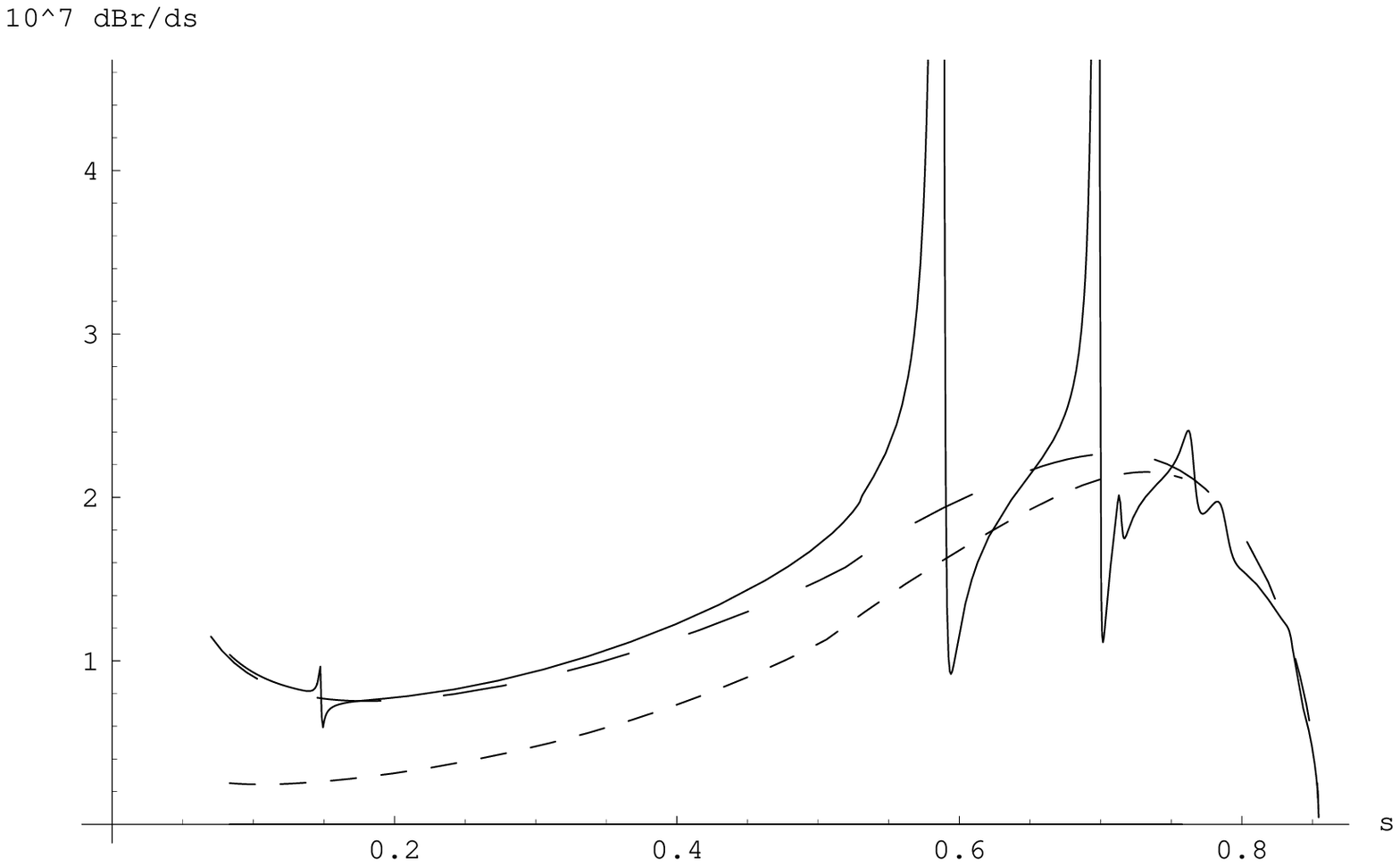}
\vskip -1.5truein
\caption[]{The same as Fig. \ref{robrbb40q2a}, but at the region 
$r_{tb} >> 1$.}
\label{robrbb40q2b}
\end{figure}

\begin{figure}[htb]
\vskip -1.5truein
\centering
\epsfxsize=3.8in
\leavevmode\epsffile{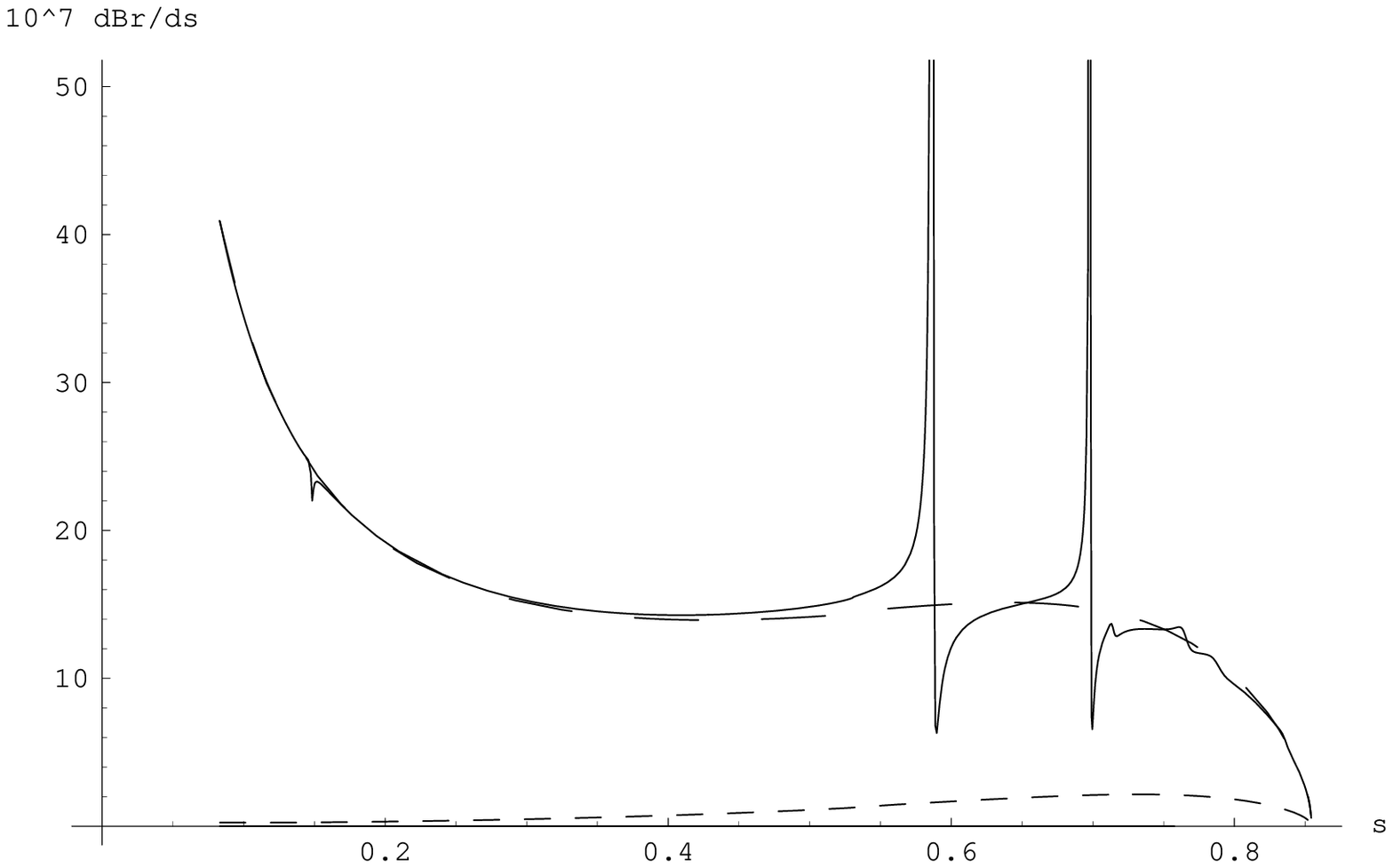}
\vskip -1.5truein
\caption[]{The same as Fig. \ref{robrbb40q2b}, 
but for fixed $\bar{\xi}_{N,bb}^{D}=90\, m_b$ value.}
\label{robrbb90q2b}
\end{figure}

In  figs.~\ref{robrbb40q2a} (\ref{robrbb40q2b}) we plot the differential 
$Br$ of the decay $\bar{B}\rightarrow  \rho e^+ e^-$ with respect to the 
dilepton mass $q^2$ for the fixed values of $\bar{\xi}_{N,bb}^{D}=40\, m_b$ 
and charged Higgs mass $m_{H^{\pm}}=400\, GeV$
at the scale $\mu=m_b$, for the ratio
$|r_{tb}|=|\frac{\bar{\xi}_{N,tt}^{U}}{\bar{\xi}_{N,bb}^{D}}| <<1$ 
($r_{tb}=\frac{\bar{\xi}_{N,tt}^{U}}{\bar{\xi}_{N,bb}^{D}} >>1$). 
The differential $Br$, obtained in the model III, is smaller
compared to the one calculated in the SM, for $|r_{tb}|<<1$.
However, it increases at the region 
$r_{tb}>> 1$ and enhances strongly compared to the SM with the 
increasing $\bar{\xi}_{N,bb}^{D}$ (Fig.~\ref{robrbb90q2b}), similar to 
the decay $\bar{B}\rightarrow \pi e^+ e^-$.
To be complete, we present the values of $Br$ for the 
$\bar{B} \rightarrow \rho e^+ e^-$ decay in the SM and model III, without the LD effects.
After integrating over $q^2$, we get
\begin{eqnarray}
Br(\bar{B} \rightarrow \rho e^+ e^-)=0.91 \times10^{-7} \,\,\, (SM)
\label{Br2SMro}
\end{eqnarray}
and for the model III
\begin{eqnarray}
Br(\bar{B} \rightarrow \rho e^+ e^-) = \left\{ \begin{array}{ll}
~ 0.44 \times 10^{-7}   & (|r_{tb}|<<1 \,,\,\bar{\xi}_{N,bb}^{D}=40\, m_b) \\ \\
~ 1.5 \times 10^{-7}   & (r_{tb}>>1 \,,\,\bar{\xi}_{N,bb}^{D}=40\, m_b ) \\ \\
~ 3.2 \times 10^{-7}   & (r_{tb}>>1 \,,\,\bar{\xi}_{N,bb}^{D}=90\, m_b)~.
\end{array} \right.
\label{Br2ro}
\end{eqnarray}
The strong enhancement of the $Br$ is observed for 
$r_{tb}>>1$, especially with increasing $\bar{\xi}_{N,bb}^{D}$.  
Note that we used the Wolfenstein parameters, $\rho=-0.07\,,\eta=0.34$, 
in the calculation of $Br$ and differential $Br$.
  
\begin{figure}[htb]
\vskip -1.5truein
\centering
\epsfxsize=3.8in
\leavevmode\epsffile{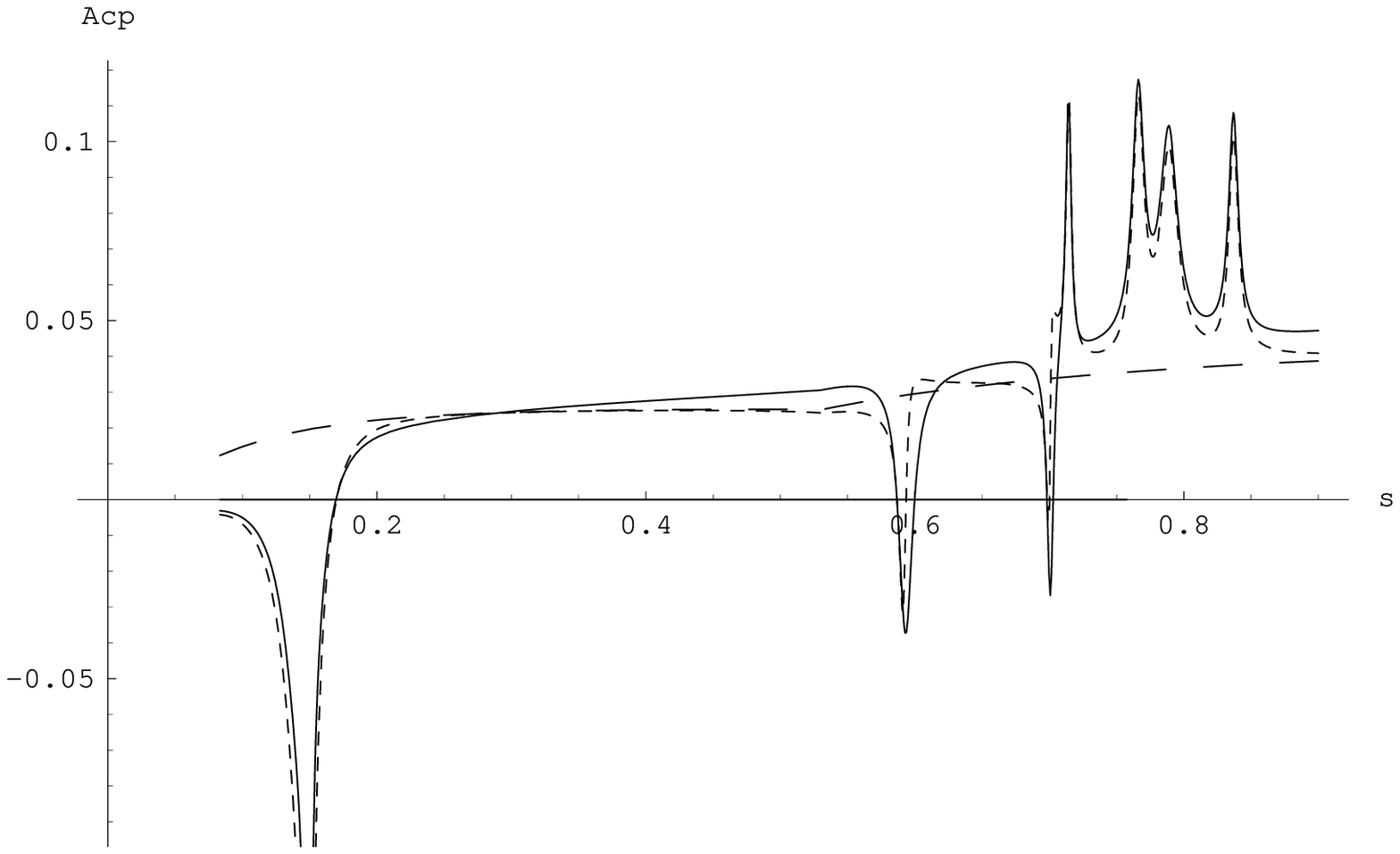}
\vskip -1.5truein
\caption[]{$A_{CP}$ as a function of  $q^2$ 
for fixed $\bar{\xi}_{N,bb}^{D}=40\, m_b$ in the region $|r_{tb}|<<1$,
at the scale $\mu=m_b$, for the process $\bar{B}\rightarrow \rho e^+ e^-$.
Here solid line corresponds to the model III with LD effects,
dashed line to the SM withouth LD effects and dotted dashed line
to the SM with LD effects.}
\label{Acprobb40q2a2}
\end{figure}

\begin{figure}[htb]
\vskip -1.5truein
\centering
\epsfxsize=3.8in
\leavevmode\epsffile{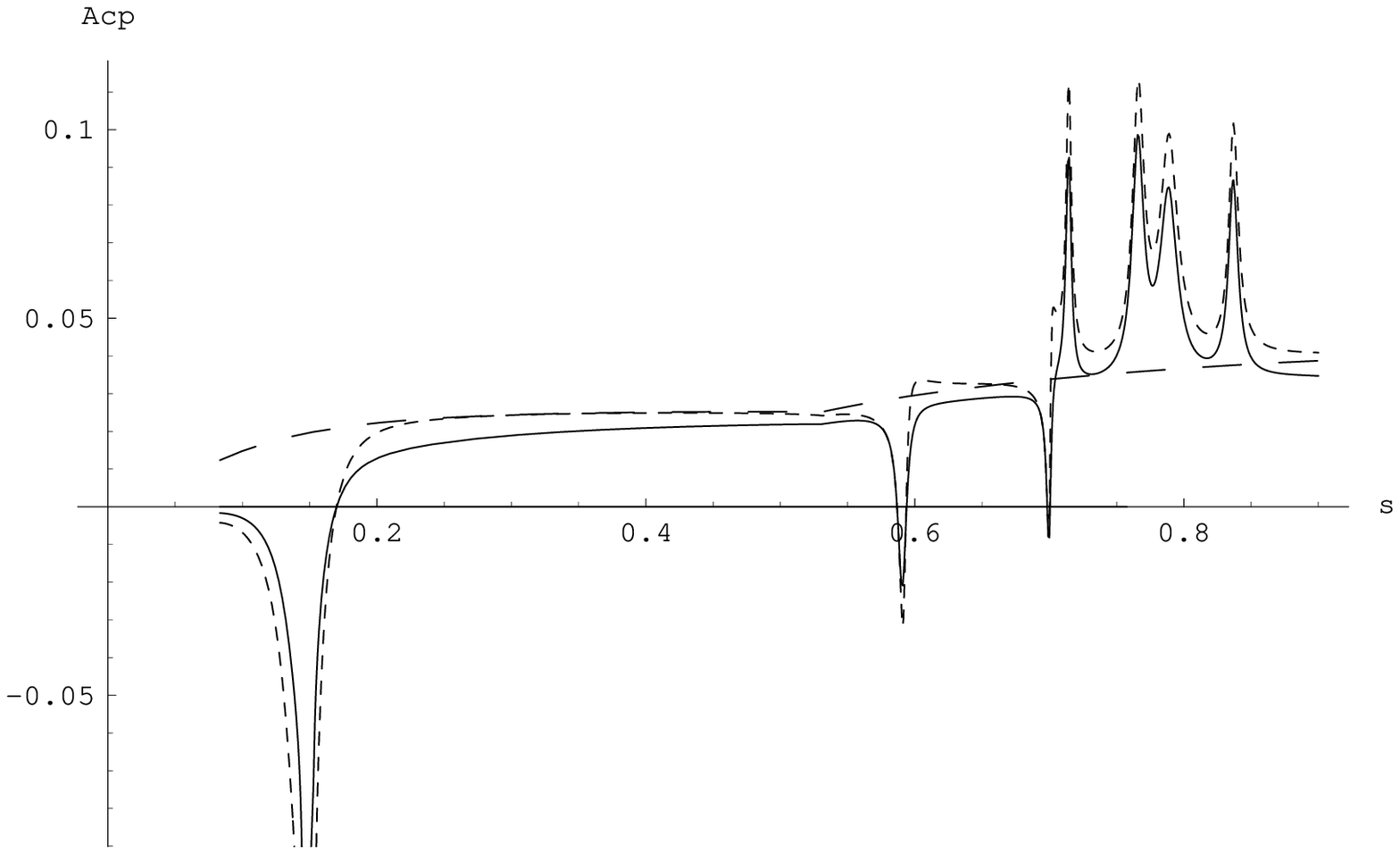}
\vskip -1.5truein
\caption[]{The same as Fig \ref{Acprobb40q2a2}, 
but at the region $r_{tb} >> 1$ .}
\label{Acprobb40q2b2}
\end{figure}

\begin{figure}[htb]
\vskip -1.5truein
\centering
\epsfxsize=3.8in
\leavevmode\epsffile{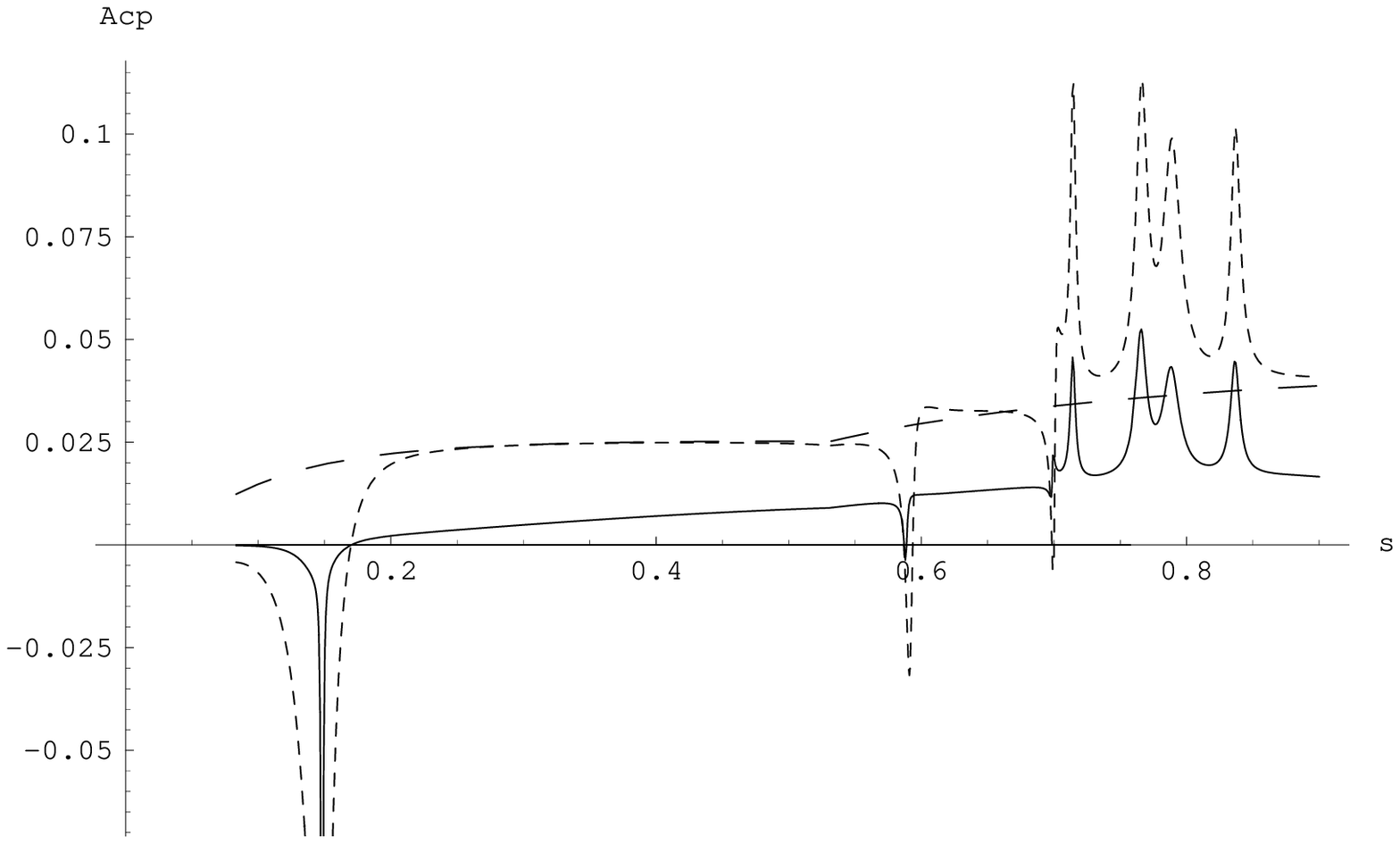}
\vskip -1.5truein
\caption[]{The same as Fig \ref{Acprobb40q2b2}, 
but for fixed $\bar{\xi}_{N,bb}^{D}=90\, m_b$ value. .}
\label{Acprobb90q2b2}
\end{figure}

Figs.~\ref{Acprobb40q2a2} and  \ref{Acprobb40q2b2} show the $q^2$ dependence 
of $A_{CP}$ for the Wolfenstein parameters $\rho=-0.07,\,\, \eta = 0.34$,
the fixed values of $\bar{\xi}_{N,bb}^{D}=40\, m_b$ 
and charged Higgs mass $m_{H^{\pm}}=400\, GeV$ at the scale $\mu=m_b$, 
for $|r_{tb}|<< 1$ and $r_{tb}>> 1$ respectively. The CP violation 
decreases in the region $r_{tb}>> 1$, especially with increasing 
$\bar{\xi}_{N,bb}^{D}$ (Fig.~\ref{Acprobb90q2b2}).
Now, we give  $<A_{CP}>$ for two different Wolfenstein parameters in two
different dilepton mass regions 
\begin{table}[h]
\small{    \begin{center}
    \begin{tabular}{|c|c|c|c|c|c|}
    \hline
    \hline \hline
    $(\rho,\eta) $   &$SM$&  model III & model III & model III \\
&    &    $\xi_{bb}^D=40 m_b$ & $\xi_{bb}^D=40\, m_b$ & $\xi_{bb}^D=90, m_b$&  \\
&    &    $r_{tb} <<1$        & $|r_{tb}|>1$          & $|r_{tb}|>1$& $q^2$ regions            \\
\hline \hline
    $(0.3, 0.34)$            &$2.00\, 10^{-2}$ &$1.90\,10^{-2}$
&$1.50\,10^{-2}$ &$0.51\, 10^{-2}$ & I  \\
    \hline
&$0.60\, 10^{-2}$ &$0.57\,10^{-2}$ &$0.53\,10^{-2}$ &$0.25\, 10^{-2}$& II \\
    \hline \hline

$(-0.07, 0.34)$ &$0.97\, 10^{-2}$ &$1.00\,10^{-2}$ 
&$0.77\,10^{-2}$ &$0.34\, 10^{-2}$& I \\
    \hline
&$0.32 \, 10^{-2}$ &$0.29\,10^{-2}$ &$0.27\,10^{-2}$ &$0.14\, 10^{-2}$& II \\
    \hline \hline    
        \end{tabular}
        \end{center} }
\caption{ The average asymmetry $<A_{CP}>$ for regions I 
( $1\, GeV\leq \sqrt q^2 \leq m_{J/\psi}-20\, MeV$ ) and II 
($m_{J/\psi}+20\, MeV \leq \sqrt q^2 \leq m_{\psi'}-20\, MeV$ ).} 
\label{Table3ro}
\end{table}

\begin{figure}[htb]
\vskip -1.5truein
\centering
\epsfxsize=3.8in
\leavevmode\epsffile{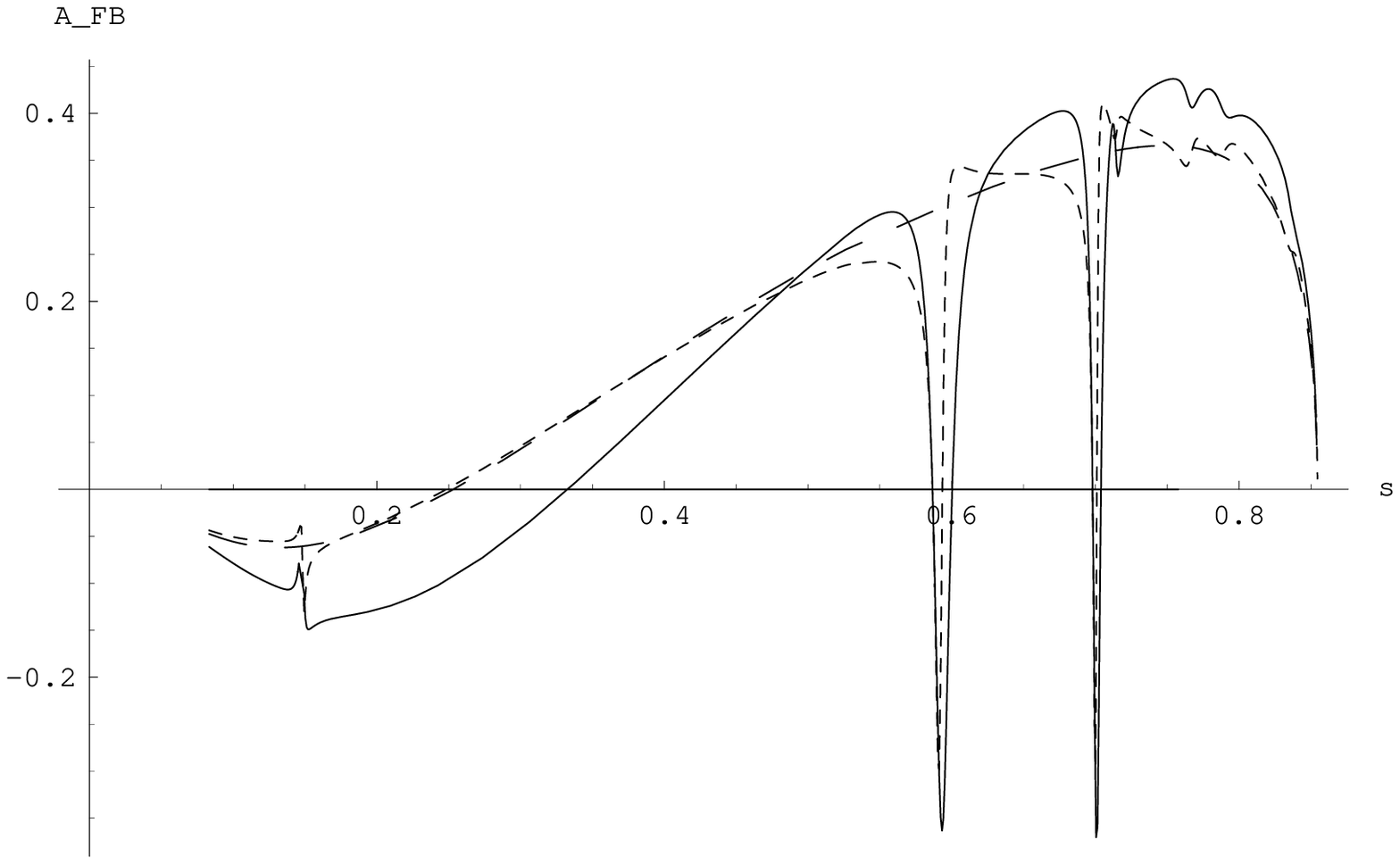}
\vskip -1.5truein
\caption[]{$A_{FB}$ as a function of  $q^2$ 
for fixed $\bar{\xi}_{N,bb}^{D}=40\, m_b$ in the region $|r_{tb}|<<1$,
at the scale $\mu=m_b$ for the process $\bar{B}\rightarrow \rho e^+ e^-$.
Here solid line and corresponds to the model III with LD effects,
dashed line to the SM withouth LD effects and dotted dashed line
to the SM with LD effects.} 
\label{AFBrobb40q2a}
\end{figure}

\begin{figure}[htb]
\vskip -1.5truein
\centering
\epsfxsize=3.8in
\leavevmode\epsffile{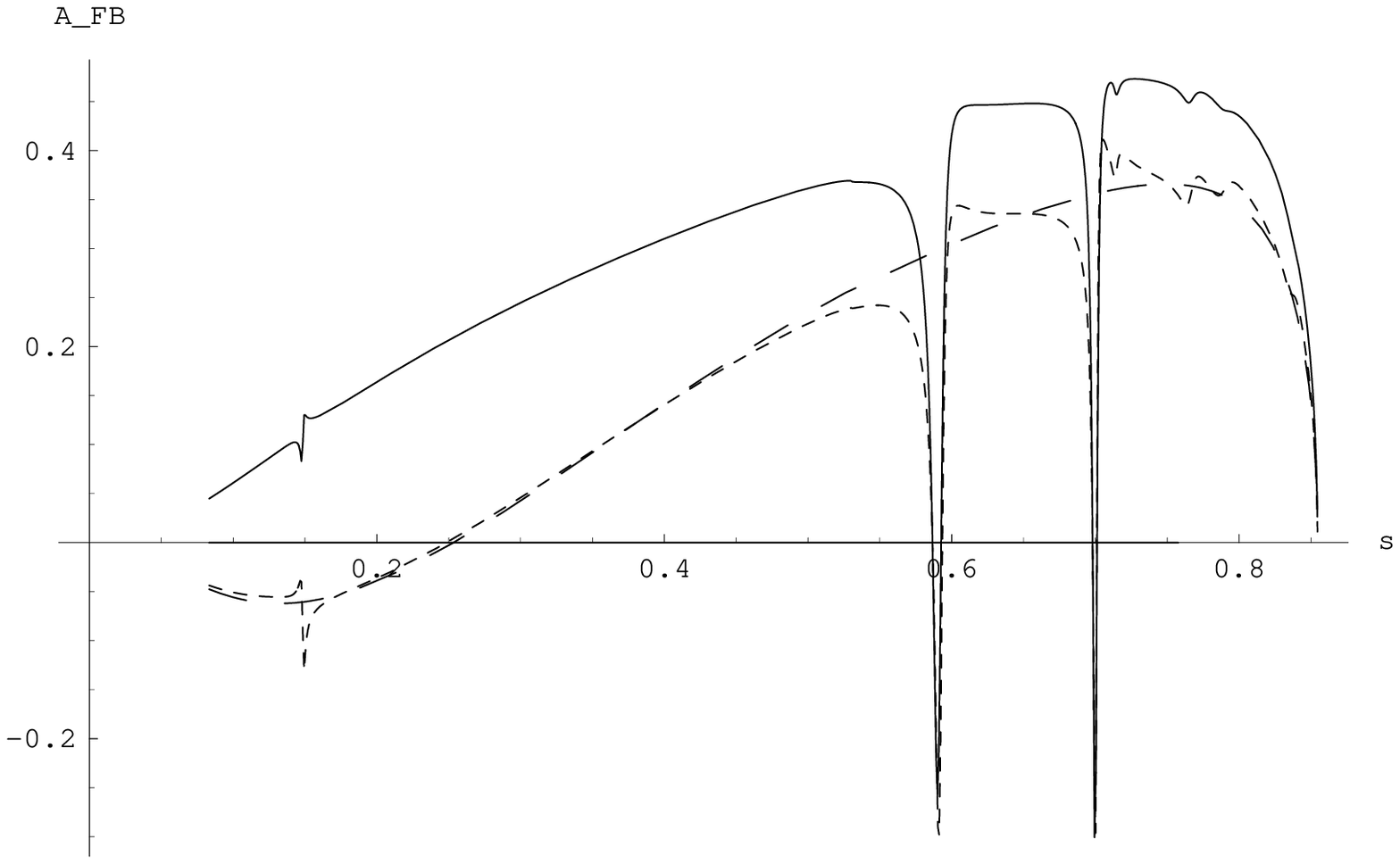}
\vskip -1.5truein
\caption[]{The same as Fig. \ref{AFBrobb40q2a}, but at the region 
$r_{tb} >> 1$.}
\label{AFBrobb40q2b}
\end{figure}

Finally, we discuss $A_{FB}$ of the process under consideration.
Figs.~\ref{AFBrobb40q2a} and  \ref{AFBrobb40q2b} show the $q^2$ dependence 
of $A_{FB}$ for the Wolfenstein parameters $\rho=-0.07,\,\, \eta = 0.34$,
the fixed values of $\bar{\xi}_{N,bb}^{D}=40\, m_b$ 
and charged Higgs mass $m_{H^{\pm}}=400\, GeV$ at the scale $\mu=m_b$, 
for $|r_{tb}|<< 1$ and $r_{tb}>> 1$ respectively.
For $r_{tb}>> 1$
(Fig.~\ref{AFBrobb40q2a}) $A_{FB}$ changes its sign almost at $s=0.34$, 
however for $r_{tb}>> 1$ (Fig.~\ref{AFBrobb40q2b}) it is positive without 
LD effects. Therefore the determination of the sign of $A_{FB}$ in the region 
$0\le s \le 0.25$ (here the upper limit corresponds to the value 
where $A_{FB}$ change sign in the SM) can give a unique information about
the existence of the model III.  

In conclusion, we analyse the selected model III parameters 
( $\bar{\xi}_{N,bb}^{D}$, $\bar{\xi}_{N,tt}^{U}$ ) 
dependencies of the differential $Br$ ,$A_{CP}$ and $A_{FB}$ of the decay
$\bar{B}\rightarrow \rho e^+ e^-$. 
We obtain that the strong enhancement of the differential $Br$ is possible 
in the framework of the model III and observe that $A_{CP}$ and $A_{FB}$ 
are very sensitive to the model III parameters 
($\bar{\xi}_{N,bb}^{D}$,  $\bar{\xi}_{N,tt}^{U}$). 

\section{Conclusion}
We study the exclusive processes 
$\bar{B}\rightarrow \pi e^+ e^-$ and $\bar{B}\rightarrow \rho e^+ e^-$ 
which are induced by the inclusive $b\rightarrow d e^+ e^-$ decay.
In such type of decays, it is informative to analyse the CP violating
effects, in addition to the quantities like $Br$, $A_{FB}$.
The origin of the CP violation in the SM is the parameter 
$\lambda_u=\frac{V_{ub}V_{ud}^*}{V_{tb}V_{td}^*}$. In the model III,
the couplings $\xi_{ij}^U$ and $\xi_{kl}^D$ 
\footnote{Here $i,j$ and $k,l$ denote up and down quarks respectively.}
can also create the CP violation in case they are not real. 
However, in our work disregard this possibility not to enlarge the number 
of free parameters and we assume that the only CP violating effect comes
from the CKM matrix elements, similar to the SM.

Now, we would like to summarize the main results of our analysis:

\begin{itemize}
\item   The $Br$ of the exclusive decays 
$\bar{B}\rightarrow \pi e^+ e^-$ and $\bar{B}\rightarrow \rho e^+ e^-$
are sensitive to the model III parameters. In the region $r_{tb}>>1$,
a strong enhancement of the $Br$ is observed with increasing 
$\bar{\xi}^D_{bb}$ in both decays eqs. As an example,
$Br_{no\, LD} (Model\,\, III) \sim  3 Br_{no \,LD}(SM)$ for 
$\bar{\xi}^D_{bb}=90\, m_b$ (\ref{Br2SM},\ref{Br2}) and 
(\ref{Br2SMro},\ref{Br2ro}). Therefore their experimental investigations 
are a crucial test for the physics beyond the SM.

\item  We calculated $<A_{CP}>$ for two different invariant mass region
(see Table (\ref{Table3}) and (\ref{Table3ro})).
We observe that $<A_{CP}>$ decreases with increasing  
$\bar{\xi}^D_{bb}$ for $r_{tb}>>1$ in both regions. 
For $r_{tb}>>1$ and $\bar{\xi}^D_{bb}=90\, m_b$, 
$<A_{CP}>$ is rather smaller compared the one 
in the SM, for both regions (region I and II), i.e.  
$<A_{CP}>_{model\, III}\,\, \sim \% 30 <A_{CP}>_{SM}$. 

\item   We calculated $A_{FB}$ for the $\bar{B}\rightarrow \rho e^+ e^-$
decay and observe that it does not change sign in the model III if the LD
effects are excluded. This shows that the determination of the sign of 
$A_{FB}$ in the region $0\leq s \leq 0.25$ will be informative to see the 
effects of the model III, if it exists.
\end{itemize}
As a conclusion, the experimental investigation of the quantities we present
here, will be an efficient tool to search for new physics beyond the SM.  
\newpage
{\bf\LARGE{Appendix}} \\

\begin{appendix}

\section{\bf The essential points of the model III.}

The Yukawa interaction in the general case of 2HDM (Model III) is
\begin{eqnarray}
{\cal{L}}_{Y}=\eta^{U}_{ij} \bar{Q}_{i L} \tilde{\phi_{1}} U_{j R}+
\eta^{D}_{ij} \bar{Q}_{i L} \phi_{1} D_{j R}+
\xi^{U}_{ij} \bar{Q}_{i L} \tilde{\phi_{2}} U_{j R}+
\xi^{D}_{ij} \bar{Q}_{i L} \phi_{2} D_{j R} + h.c. \,\,\, ,
\label{lagrangian}
\end{eqnarray}
where $L$ and $R$ denote chiral projections $L(R)=1/2(1\mp \gamma_5)$,
$\phi_{i}$ for $i=1,2$, are the two scalar doublets, $\eta^{U,D}_{ij}$
and $\xi^{U,D}_{ij}$ are the matrices of the Yukawa couplings.
The Flavor Changing (FC) part of the interaction can be written as 
\begin{eqnarray}
{\cal{L}}_{Y,FC}=
\xi^{U}_{ij} \bar{Q}_{i L} \tilde{\phi_{2}} U_{j R}+
\xi^{D}_{ij} \bar{Q}_{i L} \phi_{2} D_{j R} + h.c. \,\, ,
\label{lagrangianFC}
\end{eqnarray}
with the choice of $\phi_1$ and $\phi_2$
\begin{eqnarray}
\phi_{1}=\frac{1}{\sqrt{2}}\left[\left(\begin{array}{c c} 
0\\v+H^{0}\end{array}\right)\; + \left(\begin{array}{c c} 
\sqrt{2} \chi^{+}\\ i \chi^{0}\end{array}\right) \right]\, ; 
\phi_{2}=\frac{1}{\sqrt{2}}\left(\begin{array}{c c} 
\sqrt{2} H^{+}\\ H_1+i H_2 \end{array}\right) \,\, .
\label{choice}
\end{eqnarray}
Here the vacuum expectation values are,  
\begin{eqnarray}
<\phi_{1}>=\frac{1}{\sqrt{2}}\left(\begin{array}{c c} 
0\\v\end{array}\right) \,  \, ; 
<\phi_{2}>=0 \,\, ,
\label{choice2}
\end{eqnarray}
and the couplings  $\xi^{U,D}$ for the FC charged interactions are
\begin{eqnarray}
\xi^{U}_{ch}&=& \xi_{neutral} \,\, V_{CKM} \nonumber \,\, ,\\
\xi^{D}_{ch}&=& V_{CKM} \,\, \xi_{neutral} \,\, ,
\label{ksi1} 
\end{eqnarray}
where  $\xi^{U,D}_{neutral}$ 
\footnote{In all next discussion we denote $\xi^{U,D}_{neutral}$ 
as $\xi^{U,D}_{N}$.} 
is defined by the expression
\begin{eqnarray}
\xi^{U,D}_{N}=(V_L^{U,D})^{-1} \xi^{U,D} V_R^{U,D}\,\, .
\label{ksineut}
\end{eqnarray}
Note that the charged couplings appear as a linear combinations of neutral 
couplings multiplied by $V_{CKM}$ matrix elements (more details see
\cite{soni}). 

\section{The necessary functions appear in the calculation of the 
Wilson coefficients}

The functions $B(x)$, $C(x)$, $D(x)$, $F_{1(2)}(y)$, $G_{1(2)}(y)$, 
$H_{1}(y)$ and $L_{1}(y)$ are given as
\begin{eqnarray}
B(x)&=&\frac{1}{4}\left[\frac{-x}{x-1}+\frac{x}{(x-1)^2} \ln
x\right] \nonumber \,\, , \\
C(x)&=&\frac{x}{4}\left[\frac{x/2-3}{x-1}+\frac{3x/2+1}{(x-1)^2}
       \ln x \right] \nonumber \,\, , \\
D(x)&=&\frac{-19x^3/36+25x^2/36}{(x-1)^3}+
       \frac{-x^4/6+5x^3/3-3x^2+16x/9-4/9}{(x-1)^4}\ln x 
\nonumber \,\, ,\\
F_{1}(y)&=& \frac{y(7-5y-8y^2)}{72 (y-1)^3}+\frac{y^2 (3y-2)}{12(y-1)^4}
\,\ln y \nonumber  \,\, , \\ 
F_{2}(y)&=& \frac{y(5y-3)}{12 (y-1)^2}+\frac{y(-3y+2)}{6(y-1)^3}\, \ln y 
\nonumber  \,\, ,\\ 
G_{1}(y)&=& \frac{y(-y^2+5y+2)}{24 (y-1)^3}+\frac{-y^2} {4(y-1)^4} \, \ln y
\nonumber  \,\, ,\\ 
G_{2}(y)&=& \frac{y(y-3)}{4 (y-1)^2}+\frac{y} {2(y-1)^3} \, \ln y 
\nonumber\,\, ,\\
H_{1}(y)&=& \frac{1-4 sin^2\theta_W}{sin^2\theta_W}\,\, \frac{x
y}{8}\,\left[ 
\frac{1}{y-1}-\frac{1}{(y-1)^2} \ln y \right]\nonumber \\ 
&-& y \left[\frac{47 y^2-79 y+38}{108
(y-1)^3}-\frac{3 y^3-6 y+4}{18(y-1)^4} \ln y \right] 
\nonumber  \,\, , \\ 
L_{1}(y)&=& \frac{1}{sin^2\theta_W} \,\,\frac{x y}{8}\, \left[-\frac{1}{y-1}+
\frac{1}{(y-1)^2} \ln y \right]
\nonumber  \,\, .\\ 
\label{F1G1}
\end{eqnarray}
\section{The functions which appear in the Wilson coefficients 
$C_9^{eff}$ and $C_9^{\prime eff}$} 
The function which represents the one gluon correction to the matrix 
element $O_9$ is \cite{misiak}
\begin{eqnarray}
\tilde\eta(\hat s) = 1 + \frac{\alpha_{s}(\mu)}{\pi}\, \omega(\hat s)\,\, ,
\label{eta}
\end{eqnarray}
and
\begin{eqnarray} 
\omega(\hat s) &=& - \frac{2}{9} \pi^2 - \frac{4}{3}\mbox{Li}_2(\hat s) 
-\frac{2}{3} \ln {\hat s} \ln(1-{\hat s})-\frac{5+4{\hat s}}{3(1+2{\hat s})}
\ln(1-{\hat s}) - \nonumber \\
& &  \frac{2 {\hat s} (1+{\hat s}) (1-2{\hat s})}
{3(1-{\hat s})^2 (1+2{\hat s})} \ln {\hat s} + \frac{5+9{\hat s}-6{\hat s}^2}{6
(1-{\hat s}) (1+2{\hat s})} \,\, , 
\label{omega}
\end{eqnarray}
$h(z,\hat s)$ arises from the one loop contributions of the
four quark operators $O_1, ... ,O_6$ ($O'_1, ... ,O'_6$)
\begin{eqnarray}
h(z, \hat s) &=& -\frac{8}{9}\ln\frac{m_b}{\mu} - \frac{8}{9}\ln z +
\frac{8}{27} + \frac{4}{9} x \\
& & - \frac{2}{9} (2+x) |1-x|^{1/2} \left\{
\begin{array}{ll}
\left( \ln\left| \frac{\sqrt{1-x} + 1}{\sqrt{1-x} - 1}\right| - 
i\pi \right), &\mbox{for } x \equiv \frac{4z^2}{\hat s} < 1 \nonumber \\
2 \arctan \frac{1}{\sqrt{x-1}}, & \mbox{for } x \equiv \frac
{4z^2}{\hat s} > 1,
\end{array}
\right. \\
h(0, \hat s) &=& \frac{8}{27} -\frac{8}{9} \ln\frac{m_b}{\mu} - 
\frac{4}{9} \ln\hat s + \frac{4}{9} i\pi \,\, , 
\label{hfunc}
\end{eqnarray}
where $z=\frac{m_c}{m_b}$ and $\hat{s}=\frac{q^2}{m_b^2}$.
\end{appendix}

\newpage


\end{document}